# A comparative study of asymmetric dichotomous noise and symmetric trichotomous noise induced stochastic resonance in the globally coupled fractional oscillators

**Vishwamittar**

Department of Physics, Panjab University, Chandigarh – 160 014, India

E-mail: vm121@hotmail.com

____________________________________________________________________________

**Abstract**

The collective behaviour, in respect of stochastic resonance, has been studied in globally coupled oscillators (with fractional-order intrinsic and external damping), driven by a sinusoidal force which is either noise-free or noise-modulated, and subjected to multiplicative quadratic asymmetric dichotomous or symmetric trichotomous noise perturbing the potential parameter, the coupling factor and the local drift force. The influence of coupling between the heat bath and the applied force has been included through a simple model. The effect of variation in mass, friction and potential parameters on the output amplitude gains as function of noise-intensity, has been meticulously investigated for both types of noise and the exponents governing the dependence of collective SR peak amplitude on the three oscillator parameters have been determined and analysed. The special case arising from the zero value of the potential parameter, which implies rectilinear motion of the system particles in the absence of fluctuations, has been dealt with under the influence of the second-order asymmetric dichotomous noise and stochastic resonance has been found to occur at justifiably quite low frequencies of the external force. This brings out the importance of nonlinear term in this coloured noise, for which this phenomenon is unique. The accuracy of the analytical results has been substantiated through numerical simulations.

1. **Introduction**

Random fluctuations are inherent in the physical, chemical, and biological systems and their effects become noticeable when the size of the system is microscopic or smaller, or it is out of equilibrium. To begin with, these were believed to be detrimental and hence nuisance and, therefore, were christened noise. Depending on whether the value of the time-autocorrelation function for the model used for noise as a random process is zero or non-zero, these are classified as white noise and coloured noise. The continued thorough theoretical as well as experimental investigations have established that both these types of noise can make constructive contribution to the behaviour of various systems. It is well established that the coloured noise is more realistic and physically useful, and, as such, its realizations - dichotomous and trichotomous noises have drawn lot of attention for theoretical modelling. Though the term noise continues to be used, yet a variety of its counterintuitive effects have been well recognized and find remarkable applications in a multitude of scientific disciplines [see, e.g., 1]. One such widely studied topic, where random fluctuations are exploited in a beneficial manner, is the stochastic resonance with its different variants.

The term stochastic resonance (SR) was coined in 1981, when its underlying idea was used to suggest a probable model to describe the almost periodically recurrent ice ages on the earth. This phenomenon represents the situation where presence of noise helps in enhancing the response of a system to a weak applied temporally periodic force and makes the dependence of the output on parameters of the noise as well as of the system and the driving force to be nonmonotonic. For the initial 15 years or so, the concept continued to be essentially inconspicuous and was being studied in the nonlinear systems subjected to additive white noise and periodic signal. However, the findings that SR can play an important role in noisy biological systems and that it can be obtained through synergism between multiplicative coloured noise and linear systems, provided this a much-needed nourishment leading to its reasonably fast development. Presently, this phenomenon finds applications in numerous diverse research pursuits in physics, chemistry, biology, medical science particularly neuroscience, materials science, climatology, engineering, financial management, etc [2-10]. Interestingly, a significant amount of theoretical efforts in studying SR has been directed at investigations in the noisy oscillators – harmonic as well as bistable.

It may be mentioned that the oscillators described in terms of integer-order derivatives account for instantaneous aspects and do not take care of memory effects associated with the system as found to occur when the surroundings are colloidal glasses, dense polymer solutions, viscoelastic media, disordered semiconductors, magneto-rheological fluids, etc. In order to understand the behaviour of such systems, non-integer time-derivatives are used and the quantities corresponding to usual velocity and acceleration become fractional velocity and fractional acceleration, respectively [11,12]. The terms containing these in the equation describing the time-evolution of the system are known as fractional-order external damping and fractional-order intrinsic damping because these are associated with dissipation due to damping produced by the interaction of the oscillator with its surroundings and by the oscillator itself [13]. Both these types of damping follow power-law dependence on time. It has been convincingly proved that the physical phenomena in complex disordered materials as well as complex heterogeneous systems are explained better by the models based on fractional derivatives as compared to those involving the integer-order derivatives [14,15]. In addition to [11,12], Sun et al [16] have also summarized multifarious applications of fractional calculus.

Besides a variety of SR studies, including the effect of time–delay, pertaining to single oscillator described by integer- and fractional- order derivatives subjected to multiplicative dichotomous and trichotomous noises [see, e.g., 1, 17-31], the stochastic resonance in the collective behaviour of a large but finite number of coupled bistable systems [see, e.g., 32] as well as coupled linear oscillators [33-37] has been investigated by various workers. As far as the last category of systems is concerned, all the authors have considered the driving multiplicative noise to be linear dichotomous (nonzero mean in [33] and zero mean in all other cases) affecting the potential parameters except [37] where the friction coefficient was assumed to be fluctuating. The coupling among the oscillators has been taken to be nearest neighbour type in [33], while in the remaining four works it has been assumed to be global in the sense that every element interacts with all the other elements constituting the system. The authors in [33] also included a constant drift term that takes care of the local site interaction of the elements, which has not been considered by other workers. The oscillators were assumed to be overdamped (with friction parameter as unity) in [33-35] involving integer-order derivative in the first two and fractional-order derivative in the last. However, references [36 and 37] considered the oscillators to be underdamped having unit mass and described by the integer–order derivatives. Obviously, no work has been reported for the coupled fractional oscillators subjected to quadratic dichotomous noise or the trichotomous noise even though both are known to be more realistic and useful in describing the effect of the fluctuations [20,29]. It may be pointed out that interest in the coupled oscillators has its origin in the curiosity to look for the influence of coupling on SR and, also, in the practical applications in biological systems and complex networks [see, e.g., 38].

Incidentally, all the investigations on SR have been carried out assuming that the external periodic force interacts exclusively with the system under consideration – the oscillator, the coupled bistable oscillators and the coupled oscillators, and that the surrounding environment or the heat bath does not experience any effect. Recently, Cui and Zaccone [39], and Grabert and Thorwart [40] have highlighted this drawback in a general manner and have described some physical problems where the particles constituting the thermal bath are also driven by the applied force, which further influence the central system being studied. Both the groups have analysed the problem employing the archetypical Caldeira–Leggett model for the heat bath, by modifying the Hamiltonian to include a term corresponding to the interaction between the applied field and the central system as well as the bath oscillators. Cui and Zaccone [39] considered the classical case, derived a new generalized Langevin equation and, therefrom, the relevant classical fluctuation–dissipation theorem by assuming that the intrinsic frequency of the central oscillator is few orders of magnitude larger than that of the external periodic force. On the other hand, Grabert and Thorwart [40] treated the problem in the framework of quantum mechanics, obtained the Heisenberg equations of motion (and hence the quantum-mechanical form of the generalized Langevin equation), used this to study the problems of a damped oscillator and a dissipative two–state system, and discussed the effect on their dynamic susceptibility. However, the contribution of this effect has so far not been dealt with from the SR point-of-view.

In this contribution, we dilate upon our findings about the SR in a system of globally coupled finite number of identical oscillators described by fractional intrinsic and external damping, where both the potential parameter and the linear coupling coefficients are subjected to a random perturbation by the same multiplicative quadratic noise. Also included is an expression for the local drift force which too is under the influence of the same noise. The external periodic force is taken to be either noise-free or noise-modulated but having the same frequency. In order to examine the possible role played by the coupling of the surrounding heat bath to the applied oscillatory field in the phenomenon of SR and to develop a formalism compatible with the present work, we have dwelt upon the case of a harmonic oscillator interacting with a bath described by the independent oscillator model [41,42] in appendix A. The random fluctuations considered for the study of SR are modelled as asymmetric dichotomous noise (ADN) and symmetric trichotomous noise (STN). We have found the exact expressions for the mean displacement for the system driven by both types of noise, in the infinitely long-time limit employing the Laplace transform method and have compared their effects by studying output amplitude gains as function of noise-intensity for different values of order of fractional derivatives and correlation rates, and then by focusing our attention on dependence of the gains on variation in mass, friction, and potential parameters for typical values of order of fractional derivatives, noise correlation rate, the noise asymmetry (for ADN) and stationary probability parameter (for STN). These results have been then used to determine exponents for the variation of maximum magnitude of the gains with mass, friction and potential parameters. In fact, this aspect has not been explored in earlier works on SR. We have also delved into the situation arising from the special case of the potential parameter being zero, making the system to consist of globally coupled rectilinearly moving particles under the influence of quadratic ADN. Also included are some numerical simulation results which establish the effectiveness of the analytical method.

2. **System model and other basics**

*2.1 The model*

As a model of globally coupled oscillators, we consider a system comprising a finite number $N$ of identical elements, characterized by mass, friction, and potential parameters $m$, $\gamma$, and $k$, respectively, and are subjected to local coupling drift force $f$. All the constituent oscillators are assumed to be interacting with each other with coupling parameter $C'$, and that their dynamics is described by fractional-order intrinsic as well as external damping terms. Furthermore, the system is under the influence of a coloured noise such that the potential parameter, the drift force, and the inter-particle coupling coefficient experience the same multiplicative quadratic noise with different contributions. Besides, all these oscillators are driven by two external periodic forces having the same frequency $\Omega$, one of which is taken to be modulated by a different coloured noise.

The fractional Langevin equation (FLE) describing the dynamics of the $j$th oscillator reads

$$mD^\alpha x_j(t) + \gamma D^\beta x_j(t) + [k + a_1\chi(t) + a_2\chi^2(t)]x_j(t) = [C' + b_1\chi(t) + b_2\chi^2(t)]\sum_{i=1}^N\{x_i(t) - x_j(t)\} \\ + f[1 + e_1\chi(t) + e_2\chi^2(t)] + (1 + c)[A + B\psi(t)]\sin(\Omega t). \quad (1)$$

Here, $m$ and $\gamma$, respectively, have dimensions $MT^{\alpha-2}$ and $MT^{\beta-2}$, and

$$D^\rho x_j(t) = \frac{1}{\Gamma(n-\rho)}\int_0^t (t-t')^{(n-1-\rho)}\frac{d^n x_j(t')}{dt'^n}dt', \quad (2)$$

with $n - 1 < \rho \leq n$, $n \in \mathbb{N}$, is the Caputo time fractional derivative operator of order $\rho$ ($\rho = \alpha, \beta$), [43]. Also, $1 < \alpha \leq 2$, and $0 < \beta \leq 1$. Thus, the first two terms in Eq. (1) are expressions for intrinsic and external damping, respectively, and have power-law decay kernel. It may be mentioned that the order of the fractional derivative provides a measure of memory; a lower value implies higher memory effect and vice versa [44]. Furthermore, $\alpha = 2$ corresponds to the usual acceleration while $\beta = 1$ gives the ordinary velocity, and that for these specific values of $\alpha$ and $\beta$ (no memory effects), Eq. (1) becomes classical Langevin equation for a system of globally coupled harmonic oscillators in the presence of the random fluctuations considered here. $a_1\chi(t) + a_2\chi^2(t)$, and two other similar expressions on the right hand side in Eq. (1) represent relevant quadratic multiplicative noise $\chi(t)$ with the first term as the linear part and the second term as the nonlinear part so that the noise becomes linear if the coefficient in the latter is taken to be zero. Also, the coefficients of the linear as well as the nonlinear noise-terms in all the expressions will be assumed to be non-negative. We have taken the noise to be quadratic in nature because this is known to provide a better description of the dynamical properties of a system [45]. Also, though the noise acting on the potential parameter $k$, the coupling coefficient $C'$ and the drift force $f$, is the same $\chi(t)$, its contributions to different parameters are governed by the values of the coefficients appearing in the respective linear and non-linear parts, and, in general, these are different. Furthermore, the last term in Eq. (1), with $[A + B\psi(t)]\sin(\Omega t)$ as the applied periodic force, gives the effective force experienced by the fractional oscillator and takes into account the effect of coupling of the external sinusoidal signal to the fractional oscillator and to the surrounding thermal bath assumed to consist of $N_b$ identical harmonic oscillators, each interacting with the force with coefficients $\frac{c}{N_b}$ and having frequencies much larger than the frequency $\Omega$ of the applied force [Eq. (A12)]. Lastly, $[A + B\psi(t)]\sin(\Omega t)$ has been written as the sum of two external sinusoidal forces having same frequency, $\Omega$, and will be used as such in the derivation, but in the numerical calculations only one of these will be taken to be operative at a time. Note that in the second part, viz. $B\psi(t)\sin(\Omega t)$, the applied signal is modulated by the noise $\psi(t)$, different from the one causing fluctuations in the parameters $k$, $C'$ and $f$. Thus, amplitude $B = 0$ corresponds to the noise-free applied force, while amplitude $A = 0$ gives the noise-modulated external signal. We have ignored the additive internal noise.

## 2.2 Fractional Langevin equation for the mean field

In order to quantify the response of the system to the applied periodic signal and to investigate the SR in this, we introduce the mean field which is average of all the $N$ instantaneous positions $\{x_i(t)\}$ as

$$X(t) = \frac{\sum_{i=1}^{N} x_i(t)}{N}. \tag{3}$$

Substituting this into Eq. (1), summing over $j$ from 1 to $N$, and then dividing by $N$, we finally get

$$mD^\alpha X(t) + \gamma D^\beta X(t) + [k + a_1\chi(t) + a_2\chi^2(t)]X(t) = f[1 + e_1\chi(t) + e_2\chi^2(t)] \\ + (1+c)[A + B\psi(t)]\sin(\Omega t). \tag{4}$$

It is important to note that the expression containing the coupling parameter $C'$ and the fluctuations in this is missing in the FLE governing the mean $X(t)$, which describes the profile of average displacement of $N$ oscillators. It is this equation that will be used for discussion of SR.

However, before proceeding further, we recall that the aim of the present work is to compare the effects of ADN and STN on SR in the globally coupled double fractional–order damped oscillators. Accordingly, we first summarize the main features of these two types of noise, as per our requirement. In the derivation pertaining to ADN, the noises $\chi(t)$ and $\psi(t)$ will be denoted by $\mu(t)$ and $\xi(t)$, respectively, while $\eta(t)$ and $\zeta(t)$ will be used for these noises in the discussion for STN.

## 2.3 Asymmetric dichotomous noise

The ADN, $\mu(t)$ is assumed to be characterized by the jumps between two values $-\Delta_\mu$ and $\Delta'_\mu$ (both the noise amplitudes $\Delta_\mu$, $\Delta'_\mu > 0$) such that the rate of its transition from $-\Delta_\mu$ to $\Delta'_\mu$ is $\gamma_\mu$, while this is $\gamma'_\mu$ for the reverse jump from $\Delta'_\mu$ to $-\Delta_\mu$. Using $\Lambda_\mu$ as symbol for the noise intensity (product of the noise amplitudes), $\upsilon_\mu$ for the correlation rate and $\delta_\mu$ for the asymmetry, we have [29,46,47]

$$\Lambda_\mu = \Delta_\mu \Delta'_\mu, \quad \upsilon_\mu = \gamma_\mu + \gamma'_\mu, \text{and} \quad \delta_\mu = \Delta'_\mu - \Delta_\mu. \tag{5}$$

Also,

$$\mu^2(t) = \Lambda_\mu + \delta_\mu \mu(t), \qquad \mu^3(t) = \Lambda_\mu \delta_\mu + (\Lambda_\mu + \delta_\mu^2)\mu(t). \tag{6}$$

The stationary state statistical properties of $\mu(t)$ read

$$<\mu(t)> = 0, \qquad \text{and} \qquad <\mu(t)\mu(t')> = \Lambda_\mu \exp(-\upsilon_\mu|t - t'|). \tag{7}$$

Note that for $\delta_\mu = 0$, the noise becomes symmetric dichotomous noise (SDN) and different expressions get modified accordingly.

Replacing the symbol $\mu$ by $\xi$ in the preceding part, we get the corresponding features of the ADN $\xi(t)$. Also, the same-time cross-correlation $\Lambda_{\mu\xi}$ between the two noises, represented as

$$<\mu(t)\xi(t)> = \Lambda_{\mu\xi}, \tag{8}$$

is determined by the intensities of the two ADN and their coupling.

From Eqs. (6) and (7), we get

$$<\mu^2(t)x_j(t)> = \Lambda_\mu <x_j(t)> + \delta_\mu <\mu(t)x_j(t)>, \qquad <\mu^3(t)> = \Lambda_\mu \delta_\mu, \tag{9a}$$

and

$$<\mu^3(t)x_j(t)> = \Lambda_\mu \delta_\mu <x_j(t)> + (\Lambda_\mu + \delta_\mu^2) <\mu(t)x_j(t)>. \tag{9b}$$

Furthermore, generalization of the Shapiro-Loginov formulae of differentiation [48], for the fractional derivative of order $\rho (= \alpha, \beta)$ [22] reads

$$<\mu(t)D^\rho x_j(t)> = e^{-v_\mu t} D^\rho [<\mu(t)x_j(t)> e^{v_\mu t}]. \tag{10}$$

### 2.4 Symmetric trichotomous noise

The STN, $\eta(t)$ is a three-level random–telegraph process involving jumps between $-a_\eta, 0$ and $a_\eta$. The jumps take place according to a time-Poisson process, with switching rate $v_\eta$ and the stationary probabilities $P_s(-a_\eta) = P_s(a_\eta) = p_\eta$ and $P_s(0) = 1 - 2p_\eta$; here $0 < p_\eta \leq 1/2$ [49-51]. The mean value and the correlation function of STN are, respectively, given by

$$<\eta(t)> = 0, \quad \text{and} \quad <\eta(t)\eta(t')> = 2p_\eta a_\eta^2 \exp(-v_\eta|t - t'|). \tag{11}$$

Also, $\eta^2(t) = a_\eta^2$, the product of the two noise amplitudes, referred to as the noise intensity, will be denoted by $\sigma_\eta$ and, therefore,

$$<\eta^2(t)> = 2p_\eta \sigma_\eta. \tag{12}$$

It is pertinent to mention that for the maximum value of $p_\eta$, viz. $(p_\eta)_{\max} = 1/2$, $P_s(0) = 0$ implying that the noise switches back and forth only between the states $-a_\eta$ and $a_\eta$ with the same probability 0.5, as it happens in SDN. This, in turn, means that the STN with $p_\eta = 1/2$ has the same effect on the system as the SDN for $\Delta_\mu = a_\eta$. At the other extreme, reasonably small magnitude of $p_\eta$ indicates that the noise process $\eta(t)$ has quite low probability of occupying the nonzero states so that its average value is reasonably close to zero irrespective of magnitude of $a_\eta$. However, such a situation cannot arise in the case of SDN unless $\Delta_\mu$ and hence noise intensity $\Lambda_\mu$ is quite small. The usefulness of STN in modelling the actual noise, because of its higher flexibility, has been pointed out in many earlier works.

The relevant expressions characterizing the STN $\zeta(t)$ are obtained by replacing $\eta$ by $\zeta$. Furthermore, the same-time cross correlation between these two noises will be represented by $\sigma_{\eta\zeta}$:

$$<\eta(t)\zeta(t)> = \sigma_{\eta\zeta}. \tag{13}$$

The modified Shapiro-Loginov formulae needed for the derivations in this case are [25,27]

$$<\eta(t)D^\rho x_j(t)> = e^{-v_\eta t} D^\rho [<\eta(t)x_j(t)> e^{v_\eta t}], \tag{14}$$

and

$$<\eta^2(t)D^\rho x_j(t)> = 2p_\eta \sigma_\eta D^\rho <x_j(t)> + e^{-v_\eta t} D^\rho [e^{v_\eta t}\{<\eta^2(t)x_j(t)> - 2p_\eta \sigma_\eta <x_j(t)>\}], \tag{15}$$

for both $\rho = \alpha$ and $\beta$.

### 2.5 Digression regarding some symbols

To make the reading of the article eloquent, in this subsection, we give a brief list of some subscripts (with their meaning), which will be introduced in the next two sections.

- d     parameter pertaining to dichotomous noise;
- d, st    infinite time stationary state under the influence of dichotomous noise;
- sy    quantity corresponding to the symmetric dichotomous noise, i.e., $\delta_\mu = 0$;
- t     parameter relevant to trichotomous noise;
- t, st    stationary state in the presence of symmetric trichotomous noise;
- t, $\left(\frac{1}{2}\right)$    parameter corresponding to symmetric trichotomous noise for specific value $p_\eta = \frac{1}{2}$;

cr, $\left(\frac{1}{2}\right)$ critical value of noise intensity $\sigma_\eta$ for $p_\eta = \frac{1}{2}$.

## 3. Exact solution and output amplitude gain under the influence of ADN

The determination of collective response of the system to the ADN and the applied periodic forces, requires knowledge of the first moment of the mean displacement, $<X(t)>$. For this purpose, we average both sides of the stochastic equation, Eq. (4), over all the realizations of the trajectory using appropriate symbols for ADN and the relevant expressions from Eqs. (6) - (9), and get fractional differential equation for the time–evolution of $<X(t)>$ as

$$mD^\alpha <X(t)> + \gamma D^\beta <X(t)> + [k + a_2\Lambda_\mu]<X(t)> + [a_1 + a_2\delta_\mu]<\mu(t)X(t)> \\ = f[1 + e_2\Lambda_\mu] + (1+c)A\sin(\Omega t). \quad (16)$$

To handle the new correlator $<\mu(t)X(t)>$ in Eq. (16), we multiply both sides of Eq. (4) with $\mu(t)$, average the resulting equation, use Eqs. (6) – (10), and substitute

$$<X(t)> = y_1(t), \quad \text{and} \quad <\mu(t)X(t)> = y_2(t), \quad (17)$$

for latter convenience. This gives us

$$(a_1 + a_2\delta_\mu)\Lambda_\mu y_1(t) + e^{-v_\mu t}\{mD^\alpha + \gamma D^\beta\}\{y_2(t)e^{v_\mu t}\} + [k + a_1\delta_\mu + a_2(\Lambda_\mu + \delta_\mu^2)]y_2(t) \\ = f(e_1 + e_2\delta_\mu)\Lambda_\mu + (1+c)\Lambda_{\mu\xi}B\sin(\Omega t). \quad (18)$$

In order to solve the simultaneous equations (16) and (18), we substitute the symbols defined in Eq. (17) into the former, and use the Laplace transform method, denoting the Laplace transform of $y_n(t)$ by $Y_n(s)$, $(n = 1, 2)$. Since we are ultimately interested in the stationary state solutions corresponding to the limit $t \to \infty$, the memory effects associated with the initial conditions $y_n(0)$ and $\dot{y}_n(0)$, which appear while determining the Laplace transforms of derivatives and occur in the terms describing the now irrelevant transient state, are neglected [25,29]. Therefore, omitting the corresponding terms, we finally have

$$T_{11}Y_1(s) + T_{12}Y_2(s) = \frac{f[1+e_2\Lambda_\mu]}{s} + (1+c)A\frac{\Omega}{(s^2+\Omega^2)}, \quad (19)$$

$$T_{21}Y_1(s) + T_{22}Y_2(s) = \frac{f[e_1+e_2\delta_\mu]}{s} + (1+c)\Lambda_{\mu\xi}B\frac{\Omega}{(s^2+\Omega^2)}. \quad (20)$$

Here,

$$T_{11} = ms^\alpha + \gamma s^\beta + k + a_2\Lambda_\mu, \quad T_{12} = a_1 + a_2\delta_\mu, \quad T_{21} = \Lambda_\mu T_{12}, \quad (21a)$$

and

$$T_{22} = m(s+v_\mu)^\alpha + \gamma(s+v_\mu)^\beta + k + a_1\delta_\mu + a_2(\Lambda_\mu + \delta_\mu^2). \quad (21b)$$

We solve the simultaneous linear algebraic equations (19) and (20) using the Cramer's rule, and get

$$Y_1(s) = f[(1+e_2\Lambda_\mu)H_{d1A}(s) + (e_1 + e_2\delta_\mu)\Lambda_\mu H_{d1B}(s)]\left(\frac{1}{s}\right) \\ + (1+c)[H_{d1A}(s)A + H_{d1B}(s)\Lambda_{\mu\xi}B]\frac{\Omega}{(s^2+\Omega^2)}, \quad (22)$$

where

$$H_{d1A}(s) = \frac{T_{22}}{D_d(s)}, \quad \text{and} \quad H_{d1B}(s) = -\frac{T_{12}}{D_d(s)}; \quad (23)$$

with

$$D_d(s) = T_{11}T_{22} - T_{12}T_{21}. \quad (24)$$

We have ignored expression for $Y_2(s)$ as this is not needed in further discussion. It may be mentioned that for the special case $\delta_\mu = 0$, i.e, for the SDN,

$$H_{d1A}(s, \delta_\mu = 0) = \frac{m(s+\nu_\mu)^\alpha + \gamma(s+\nu_\mu)^\beta + k + a_2\Lambda_\mu}{D_{sy}(s)}, \quad \text{and} \quad H_{d1B}(s, \delta_\mu = 0) = -\frac{a_1}{D_{sy}(s)}, \quad (25)$$

with

$$D_{sy}(s) = \{ms^\alpha + \gamma s^\beta + k + a_2\Lambda_\mu\}\{m(s+\nu_\mu)^\alpha + \gamma(s+\nu_\mu)^\beta + k + a_2\Lambda_\mu\} - a_1^2\Lambda_\mu. \quad (26)$$

Furthermore, finding the inverse Laplace transform of $Y_1(s)$, we obtain expression for $y_1(t) = <X(t)>$ in the infinite-time limit, which defines the stationary state of the system. Thus, when the effects of initial conditions have subsided and only the drift force and the applied periodic force of frequency $\Omega$ are efficacious, the average displacement of the collection of oscillators is given by

$$\begin{aligned} y_1(t)|_{t \to \infty} &= <X(t)>|_{d,t \to \infty} \equiv <X(t)>_{d,st} \\ &= f(1 + e_2\Lambda_\mu)\int_0^t h_{d1A}(t - t')\, dt' + f(e_1 + e_2\delta_\mu)\Lambda_\mu \int_0^t h_{d1B}(t - t')\, dt' \\ &+ (1 + c)A \int_0^t h_{d1A}(t - t')\sin(\Omega t')\, dt' + (1 + c)\Lambda_{\mu\xi}B \int_0^t h_{d1B}(t - t')\sin(\Omega t')\, dt'. \quad (27) \end{aligned}$$

It may be noted that $h_{d1A}(t - t')$ and $h_{d1B}(t - t')$ represent the retarded response (in the form of displacement) of the system, and, therefore, $h_{d1A}(t)$ and $h_{d1B}(t)$, which are inverse Laplace transforms of $H_{d1A}(s)$ and $H_{d1B}(s)$, respectively, are referred to as relaxation functions at time $t$. In the light of linear response theory [25], we write Eq. (27) as

$$<X(t)>_{d,st} = f[(1 + e_2\Lambda_\mu)H_{d1A}(s = 0) + (e_1 + e_2\delta_\mu)\Lambda_\mu H_{d1B}(s = 0)]$$

$$+ (1 + c)[A_{d1}\sin(\Omega t + \theta_{d1A}) + \Lambda_{\mu\xi}B_{d1}\sin(\Omega t + \theta_{d1B})]. \quad (28)$$

Here, the time–independent terms involving $f$ arise from the contribution of the local fluctuating drift force, while the sinusoidal terms with $(1 + c)$ as multiplier account for the effect of the applied periodic signals, $A\sin(\Omega t)$ and $B\xi(t)\sin(\Omega t)$, with amplitudes $(1 + c)A_{d1}$ and $(1 + c)\Lambda_{\mu\xi}B_{d1}$, and phase shifts $\theta_{d1A}$ and $\theta_{d1B}$, respectively, with

$$A_{d1} = A|H_{d1A}(s = -i\Omega)| = A\left[\frac{N_{d1A1}^2 + N_{d1A2}^2}{D_{d1}^2 + D_{d2}^2}\right]^{1/2}, \quad (29a)$$

$$\Lambda_{\mu\xi}B_{d1} = \Lambda_{\mu\xi}B|H_{d1B}(s = -i\Omega)| = \Lambda_{\mu\xi}B\left[\frac{N_{d1B1}^2 + N_{d1B2}^2}{D_{d1}^2 + D_{d2}^2}\right]^{1/2}; \quad (29b)$$

$$\theta_{d1A} = \arctan\left[\frac{N_{d1A1}D_{d2} - N_{d1A2}D_{d1}}{N_{d1A1}D_{d1} + N_{d1A2}D_{d2}}\right], \quad \text{and} \quad \theta_{d1B} = \arctan\left[\frac{N_{d1B1}D_{d2} - N_{d1B2}D_{d1}}{N_{d1B1}D_{d1} + N_{d1B2}D_{d2}}\right]. \quad (30)$$

Note that $N_{d1A1}$ and $N_{d1A2}$ are the real and imaginary parts of the numerator of $H_{d1A}(s = -i\Omega)$, while $D_{d1}$ and $D_{d2}$ are similar quantities pertaining to its denominator, $D_d(s = -i\Omega)$. Replacing $A$ by $B$, we get the corresponding terms $N_{d1B1}$ and $N_{d1B2}$ in the preceding equations. Also, the indirect effect having its origin in the interaction of the applied sinusoidal force with the heat bath is reflected in the presence of $c$ in Eq. (28).

It is pertinent to note that the drift force $f$ and its fluctuations, effectively lead to the positional shift in $<X(t)>_{d,st}$ as their contribution is time–independent; and that both of these terms will be present whether $B = 0$ or $A = 0$. However, since these terms do not affect the behavioural dependence of the response of the system to the applied periodic force, we ignore these in the discussion of SR. Furthermore, as the factor $(1 + c)$ is constant for the heat bath considered here, we define the relevant stationary state output amplitude gains as ratio between the amplitude of the system output and the product of $(1 + c)$ and the amplitude of the corresponding applied signal:

$$GA_{d1} = (1 + c)A_{d1}/[(1 + c)A], \; (B = 0); \quad \text{and} \quad GB_{d1} = (1 + c)\Lambda_{\mu\xi}B_{d1}/[(1 + c)B], \; (A = 0). \quad (31)$$

It may be pointed out that for the solution (27), and, hence Eq. (28) to be stable and, therefore, causal, the noise intensity $\Lambda_\mu$ should be not only positive but also be less than a critical value, $(\Lambda_\mu)_{cr}$, which is the smallest positive root of $D_d(s = 0) = 0$ [1,52]. Thus, the stability criterion reads

$$0 < \Lambda_\mu < (\Lambda_\mu)_{cr}, \quad (32)$$

where for $a_2 = 0$,

$$(\Lambda_\mu(a_2 = 0))_{\text{cr}} = \frac{(m\nu_\mu^\alpha + \gamma\nu_\mu^\beta + k + a_1\delta_\mu)k}{a_1^2} ; \quad (33)$$

and for $a_2 \neq 0$,

$$(\Lambda_\mu)_{\text{cr}} = \frac{-B_\mu - \{B_\mu^2 - 4a_2^2 C_\mu\}^{1/2}}{2a_2^2}, \quad (34)$$

with

$$B_\mu = (m\nu_\mu^\alpha + \gamma\nu_\mu^\beta + 2k)a_2 - (a_1 + a_2\delta_\mu)a_1, \quad \text{and} \quad C_\mu = [(m\nu_\mu^\alpha + \gamma\nu_\mu^\beta + k) + (a_1 + a_2\delta_\mu)\delta_\mu]k. \quad (35)$$

Eqs. (32) and (34) show that when the system is under the influence of quadratic ADN, different parameters must be so chosen that $B_\mu < 0$ and $B_\mu^2 \geq 4a_2^2 C_\mu$. This, in general, requires that $a_2$ be much smaller than $a_1$. Furthermore, for SDN, $(\Lambda_\mu)_{\text{cr}}$ can be obtained by substituting $\delta_\mu = 0$ in the preceding relevant expressions. As a special case of this, if we substitute $m = 0, \gamma = 1, a_1 = 1, a_2 = b_1 = b_2 = 0$, then we get from Eq. (33)

$$(\Lambda_\mu)_{\text{cr}} = k(k + \nu_\mu^\beta), \quad (36)$$

as found by Zhong et al [36].

A look at Eq. (16), the fractional differential equation describing the average displacement of the mean field, viz., $<X(t)> = y_1(t)$, shows that presence of the nonlinear term ($a_2 > 0$) in the noise increases the value of the potential parameter from $k$ to $(k + a_2\Lambda_\mu)$. Moreover, this noise–modulated effective potential parameter does not vanish even when $k = 0$ which, in fact, implies rectilinear motion of the system. This is an interesting situation as the second-order term in the multiplicative noise associated with $X(t)$ makes the rectilinear motion of the collection to be oscillatory with frequency proportional to $(a_2\Lambda_\mu)^{1/2}$ and this is expected to produce SR. This, indeed, is found to be so. In this case, the critical value of the noise intensity is given by

$$(\Lambda_\mu)_{\text{cr}} = -\frac{(m\nu_\mu^\alpha + \gamma\nu_\mu^\beta)a_2 - (a_1 + a_2\delta_\mu)a_1}{a_2^2}. \quad (37)$$

Obviously, for $k = 0$, $(\Lambda_\mu)_{\text{cr}}$ will be positive only if $a_2 < \frac{(a_1 + a_2\delta_\mu)}{(m\nu_\mu^\alpha + \gamma\nu_\mu^\beta)} a_1$, which will be so if $a_1$ is significantly larger than $a_2$ unless $(m\nu_\mu^\alpha + \gamma\nu_\mu^\beta)$ is very small.

Next, denoting the deviation of the $j$th particle's instantaneous displacement from the value of the mean field at that time by $\epsilon_j(t)$, we have

$$x_j(t) = X(t) + \epsilon_j(t). \quad (38)$$

Substituting this into Eq. (1) together with Eq. (3), using Eq. (4) with noise symbols as for ADN and proceeding as in [34-37], we also finally find that in the infinite–time limit the mean deviations are given by

$$<\epsilon_j(t)>_{\text{st}} = 0; \quad (39)$$

for $j = 1, 2, \ldots, N$. This brings out the synchronization of the trajectories of the individual oscillators with that of the mean field in the steady state. The inequalities giving the stability criterion of this synchronization behaviour are obtained from Eqs. (33) – (35) by making the following replacements

$$k \to k + NC', \quad a_1 \to a_1 + Nb_1, \quad \text{and} \quad a_2 \to a_2 + Nb_2. \quad (40)$$

However, we do not dwell upon this aspect much as the findings of [34-37] will be valid here as well. Rather, we focus our effort on discussing phenomenon of SR in the collective system by analysing the dependence of $GA_{d1}$ and $GB_{d1}$, on various parameters in section 5.

## 4. Exact solution and output amplitude gain for STN

In order to investigate the collective response of the system under consideration and subjected to STN, we use the symbols for this noise in Eq. (4) and implement the usual procedure comprising the following steps:

(i) average both sides of the relevant fractional differential equation over all realizations of the trajectory and simplify this employing Eqs. (11) - (13);

(ii) take care of the correlators $<\eta(t)X(t)>$ and $<\eta^2(t)X(t)>$ appearing in the equation got in step (i), by successively multiplying both sides of Eq. (4) with $\eta(t)$ as well as $\eta^2(t)$, averaging the two fractional differential equations so obtained and simplifying these using Eqs. (11) - (15);

(iii) substitute $<X(t)> = z_1(t)$, $<\eta(t)X(t)> = z_2(t)$, and $<\eta^2(t)X(t)> = z_3(t)$, in the three equations obtained in the preceding two steps.

Thus, we finally get the following fractional differential equations

$$[mD^\alpha + \gamma D^\beta + k]z_1(t) + a_1 z_2(t) + a_2 z_3(t) = f[1 + 2e_2 p_\eta \sigma_\eta] + (1+c) A \sin(\Omega t), \tag{41}$$

$$(k + a_2\sigma_\eta)z_2(t) + e^{-v_\eta t}\{mD^\alpha + \gamma D^\beta\}\{z_2(t)e^{v_\eta t}\} + a_1 z_3(t) = 2f e_1 p_\eta \sigma_\eta + (1+c)\sigma_{\eta\zeta} B \sin(\Omega t), \tag{42}$$

and

$$2p_\eta\sigma_\eta\{mD^\alpha + \gamma D^\beta\} z_1(t) - 2p_\eta\sigma_\eta e^{-v_\eta t}\{mD^\alpha + \gamma D^\beta\}\{z_1(t)e^{v_\eta t}\} + a_1\sigma_\eta z_2(t) + (k + a_2\sigma_\eta)z_3(t)$$
$$+ e^{-v_\eta t}\{mD^\alpha + \gamma D^\beta\}\{z_3(t)e^{v_\eta t}\} = 2p_\eta\sigma_\eta[f(1 + e_2\sigma_\eta) + (1+c) A \sin(\Omega t)]. \tag{43}$$

To simplify the equations appearing in further derivation, we multiply both the sides of Eq. (41) with $2p_\eta\sigma_\eta$, and subtract this from Eq. (43). This gives us

$$-2p_\eta\sigma_\eta k z_1(t) - 2p_\eta\sigma_\eta e^{-v_\eta t}\{mD^\alpha + \gamma D^\beta\}\{z_1(t)e^{v_\eta t}\} + (1 - 2p_\eta)a_1\sigma_\eta z_2(t)$$
$$+\{k + (1 - 2p_\eta) a_2\sigma_\eta\}z_3(t) + e^{-v_\eta t}\{mD^\alpha + \gamma D^\beta\}\{z_3(t)e^{v_\eta t}\} = 2f(1 - 2p_\eta)e_2 p_\eta \sigma_\eta^2. \tag{44}$$

In order to determine the stationary state solutions ($t \to \infty$), we find Laplace transforms of different terms on both the sides in Eqs. (41), (42) and (44), using $Z_n(s)$ for the Laplace transform of $z_n(t)$, ($n = 1, 2, 3$). Thus, we finally get

$$S_{11}Z_1(s) + S_{12}Z_2(s) + S_{13}Z_3(s) = \frac{f[1 + 2e_2 p_\eta \sigma_\eta]}{s} + (1+c) A \frac{\Omega}{(s^2 + \Omega^2)}, \tag{45}$$

$$0 \, Z_1(s) + S_{22}Z_2(s) + S_{23}Z_3(s) = \frac{2f e_1 p_\eta \sigma_\eta}{s} + (1+c)\sigma_{\eta\zeta} B \frac{\Omega}{(s^2 + \Omega^2)}, \tag{46}$$

$$S_{31}Z_1(s) + S_{32}Z_2(s) + S_{33}Z_3(s) = \frac{2f(1 - 2p_\eta)e_2 p_\eta \sigma_\eta^2}{s}. \tag{47}$$

The coefficients $\{S_{ln}\}$, with $l, n = 1, 2, 3$ are given by

$$S_{11} = ms^\alpha + \gamma s^\beta + k, \quad S_{12} = S_{23} = a_1, \quad S_{13} = a_2, \quad S_{22} = m(s + v_\eta)^\alpha + \gamma(s + v_\eta)^\beta + k + a_2\sigma_\eta,$$

$$S_{31} = -2p_\eta\sigma_\eta[m(s + v_\eta)^\alpha + \gamma(s + v_\eta)^\beta + k], \quad S_{32} = (1 - 2p_\eta) a_1\sigma_\eta, \tag{48}$$

$$S_{33} = m(s + v_\eta)^\alpha + \gamma(s + v_\eta)^\beta + k + (1 - 2p_\eta) a_2\sigma_\eta.$$

Solving Eqs. (45) - (47), we get for $Z_1(s)$, which is the only quantity of interest in the discussion of SR,

$$Z_1(s) = f[(1 + 2e_2 p_\eta \sigma_\eta) H_{t1A}(s) + 2e_1 p_\eta \sigma_\eta H_{t1B}(s) - 2(1 - 2p_\eta)e_2 p_\eta \sigma_\eta^2 H_{t1C}(s)]\left(\frac{1}{s}\right)$$
$$+ (1+c)[H_{t1A}(s)A + H_{t1B}(s)\sigma_{\eta\zeta} B] \frac{\Omega}{(s^2 + \Omega^2)}. \tag{49}$$

Here,

$$H_{t1A}(s) = \frac{S_{22}S_{33} - S_{23}S_{32}}{D_t(s)}, \qquad H_{t1B}(s) = \frac{S_{13}S_{32} - S_{12}S_{33}}{D_t(s)}, \qquad H_{t1C}(s) = \frac{S_{13}S_{22} - S_{12}S_{23}}{D_t(s)}, \tag{50}$$

and

$$D_t(s) = S_{11}(S_{22}S_{33} - S_{23}S_{32}) + S_{31}(S_{12}S_{23} - S_{13}S_{22}). \tag{51}$$

Before proceeding with the general aspects, we look at some expressions of further interest, obtained from Eqs. (50) and (51) for $p_\eta = 1/2$. In this case, $D_t(s)$ is given by

$$D_{t,\left(\frac{1}{2}\right)}(s) = \{m(s + \nu_\eta)^\alpha + \gamma(s + \nu_\eta)^\beta + k\} D'_{t,\left(\frac{1}{2}\right)}(s), \tag{52}$$

and $H_{t1A}(s), H_{t1B}(s)$ by

$$H_{t1A,\left(\frac{1}{2}\right)}(s) = \frac{m(s+\nu_\eta)^\alpha + \gamma(s+\nu_\eta)^\beta + k + a_2\sigma_\eta}{D'_{t,\left(\frac{1}{2}\right)}(s)}, \qquad H_{t1B,\left(\frac{1}{2}\right)}(s) = -\frac{a_1}{D'_{t,\left(\frac{1}{2}\right)}(s)}, \tag{53}$$

where

$$D'_{t,\left(\frac{1}{2}\right)}(s) = (ms^\alpha + \gamma s^\beta + k + a_2\sigma_\eta)\{m(s+\nu_\eta)^\alpha + \gamma(s+\nu_\eta)^\beta + k + a_2\sigma_\eta\} - a_1^2\sigma_\eta. \tag{54}$$

Using the correspondence $\Lambda_\mu \leftrightarrow \sigma_\eta$, $\nu_\mu \leftrightarrow \nu_\eta$, and comparing Eqs. (25) and (26) with Eqs. (52) and (54), we find that (i) $D_{sy}(s)$ differs from $D_{t,\left(\frac{1}{2}\right)}(s)$ but it is identical with $D'_{t,\left(\frac{1}{2}\right)}(s)$; (ii) $H_{d1A}(s, \delta_\mu = 0)$ and $H_{d1B}(s, \delta_\mu = 0)$ are the same as $H_{t1A,\left(\frac{1}{2}\right)}(s)$ and $H_{t1B,\left(\frac{1}{2}\right)}(s)$, respectively. The observation (ii) is in conformity with the fact that for $p_\eta = 1/2$, STN reduces to SDN.

Inverse Laplace transform of both the sides of Eq. (49) gives us $< X(t) >$ in the $t \to \infty$ limit, having three time–independent $f$ containing terms and two frequency– and time– dependent terms involving $(1 + c)$. As explained in section 3, we omit the terms involving $f$ and, therefore, write the final solution representing the stationary state response of the system to the applied sinusoidal signals and under the influence of STN as

$$< X(t) >_{t,st} = (1 + c)[A_{t1}\sin(\Omega t + \theta_{t1A}) + \sigma_{\eta\zeta}B_{t1}\sin(\Omega t + \theta_{t1B})]. \tag{55}$$

Here, $(1 + c)A_{t1}$ and $(1 + c)\sigma_{\eta\zeta}B_{t1}$, are the response amplitudes, while $\theta_{t1A}$ and $\theta_{t1B}$ are the phase shifts associated with the forces $A \sin(\Omega t)$ and $B\zeta(t)\sin(\Omega t)$, such that

$$A_{t1} = A|H_{t1A}(s = -i\Omega)| = A\left[\frac{N_{t1A1}^2 + N_{t1A2}^2}{D_{t1}^2 + D_{t2}^2}\right]^{1/2}, \tag{56a}$$

$$\sigma_{\eta\zeta}B_{t1} = \sigma_{\eta\zeta}B|H_{t1B}(s = -i\Omega)| = \sigma_{\eta\zeta}B\left[\frac{N_{t1B1}^2 + N_{t1B2}^2}{D_{t1}^2 + D_{t2}^2}\right]^{1/2}; \tag{56b}$$

$$\theta_{t1A} = \arctan\left[\frac{N_{t1A1}D_{t2} - N_{t1A2}D_{t1}}{N_{t1A1}D_{t1} + N_{t1A2}D_{t2}}\right], \qquad \text{and} \qquad \theta_{t1B} = \arctan\left[\frac{N_{t1B1}D_{t2} - N_{t1B2}D_{t1}}{N_{t1B1}D_{t1} + N_{t1B2}D_{t2}}\right]. \tag{57}$$

The symbols $N_{t1A1}$ and $N_{t1A2}$, used in Eqs. (56a) and (57), stand for real and imaginary parts of the numerator of $H_{t1A}(s = -i\Omega)$, and $D_{t1}$ and $D_{t2}$ are the corresponding quantities for the denominator, $D_t(s = -i\Omega)$. Replacing $A$ by $B$, we have similar terms appearing in Eqs. (56b) and (57). The stationary state output amplitude gains for this case are defined by

$$GA_{t1} = (1 + c)A_{t1}/[(1 + c)A], \ (B = 0); \quad \text{and} \quad GB_{t1} = (1 + c)\sigma_{\eta\zeta}B_{t1}/[(1 + c)B], \ (A = 0). \tag{58}$$

The condition for the stability of the solution is

$$0 < \sigma_\eta < (\sigma_\eta)_{cr}. \tag{59}$$

For $a_2 = 0$ the critical noise intensity, $(\sigma_\eta)_{cr}$, is given by

$$(\sigma_\eta(a_2 = 0))_{\text{cr}} = \frac{kR_1^2}{a_1^2 R_2}; \tag{60}$$

with

$$R_1 = (mv_\eta^\alpha + \gamma v_\eta^\beta + k), \quad \text{and} \quad R_2 = (1 - 2p_\eta)k + 2p_\eta R_1. \tag{61}$$

On the other hand, when $a_2 \neq 0$,

$$(\sigma_\eta)_{\text{cr}} = \frac{-B_\eta - \{B_\eta^2 - 4ka_2^2 R_1^2 R_2\}^{1/2}}{2a_2^2 R_2}, \tag{62}$$

where

$$B_\eta = -a_1^2 R_2 + a_2 R_1(k + R_2). \tag{63}$$

For the specific value $p_\eta = 1/2$, $R_2 = R_1$, so that when $a_2 = 0$, Eq. (60) becomes

$$(\sigma_\eta(a_2 = 0))_{\text{cr},\left(\frac{1}{2}\right)} = \frac{kR_1}{a_1^2} = \frac{k(mv_\eta^\alpha + \gamma v_\eta^\beta + k)}{a_1^2}; \tag{64}$$

while for $a_2 \neq 0$, Eq. (62) reduces to

$$(\sigma_\eta)_{\text{cr},\left(\frac{1}{2}\right)} = \frac{-[-a_1^2 + a_2(k+R_1)] - \{[-a_1^2 + a_2(k+R_1)]^2 - 4ka_2^2 R_1\}^{1/2}}{2a_2^2}. \tag{65}$$

As expected, the expressions in Eqs. (64) and (65) are identical with the corresponding results for SDN; viz., Eqs. (33) – (35) with $\delta_\mu = 0$. Of course, in general, $(\sigma_\eta)_{\text{cr}}$ and $(\Lambda_\mu)_{\text{cr}}$ are different from each other.

It is important to point out that if we substitute $k = 0$ in Eq. (41), the oscillatory term vanishes; this is in contrast with the observation made after equation (36) describing the case when the system is subjected to quadratic dichotomous noise – ADN as well as SDN. Thus, we shall not get SR for $k = 0$ in the system driven by STN. Nonetheless, the remarks regarding synchronization behaviour made in the last part of section 3, including the correspondence mentioned in Eq. (40), hold good in this case also.

## 5. Numerical results and discussion

With a view to comprehend the effect of different parameters on the output amplitude gains of the mean field for the globally coupled oscillators driven by ADN and STN considered here, we have first found $(\Lambda_\mu)_{\text{cr}}$ and $(\sigma_\eta)_{\text{cr}}$ using the relevant equations derived in sections 3 and 4, for the chosen set of the parameter values. These values have been then used as upper limit for the range of $\Lambda_\mu$ and $\sigma_\eta$ values to make sure that the stability inequality conditions in Eqs. (32) and (59), and the related ones, are satisfied while obtaining the plots of the gains $GA_{\text{d1}}$, $GB_{\text{d1}}$ and $GA_{\text{t1}}$, $GB_{\text{t1}}$, for that particular case. Furthermore, in order to keep the effect of the non-modulated and the modulated applied periodic forces in proper perspective, both $\Lambda_{\mu\xi}$ and $\sigma_{\eta\zeta}$ have been taken to be 1.0 in all the numerical calculations.

From the expressions for $(\Lambda_\mu)_{\text{cr}}$ and $(\sigma_\eta)_{\text{cr}}$ it is clear that for a specific choice of other parameters, the critical noise intensities increase monotonically with increase in the values of $m$, $\gamma$ and $k$; this, in turn, implies higher range of allowed noise intensities, which, indeed, has been found to be true.

It is interesting to note that if we assume the noise modulating the applied force, $B \sin(\Omega t)$, to be the same as the one perturbing other parameters, then $\Lambda_{\mu\xi}$ in Eq. (18) and $\sigma_{\eta\zeta}$ in Eq. (42) get replaced by $\Lambda_\mu$ and $\sigma_\eta$, respectively. This ultimately modifies the expressions (31b) and (58b) for the gains $GB_{\text{d1}}$ and $GB_{\text{t1}}$ to read

$$GB_{\text{d1}} = \Lambda_\mu B_{\text{d1}}/B, \quad \text{and} \quad GB_{\text{t1}} = \sigma_\eta B_{\text{t1}}/B. \tag{66}$$

The presence of the noise intensities as multipliers in place of fixed cross-correlations severely affects the variation of these two gains as function of $\Lambda_\mu$ and $\sigma_\eta$, respectively, and hence the nature of the relevant plots. In fact, it is observed that in most of the cases, these curves exhibit monotonous increase rather than a SR peak for the noise intensities satisfying Eqs (32) and (59).

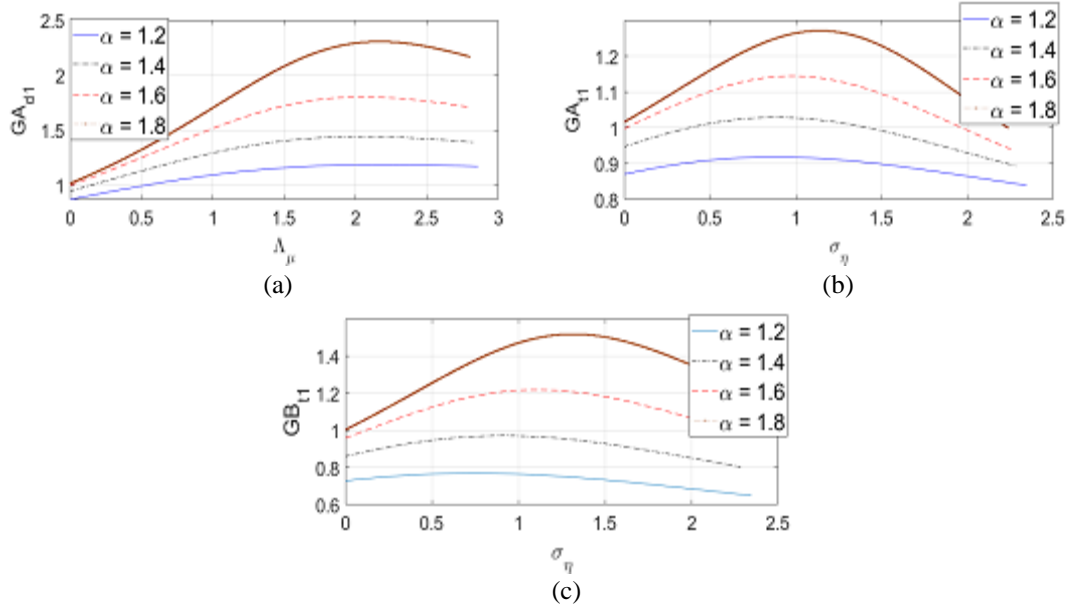

**Fig. 1.** Dependence of gains vs noise intensity on $\alpha$ : (a) $GA_{d1}$, (b) $GA_{t1}$, and (c) $GB_{t1}$ for $m = 1.0, \gamma = 0.4, k = 1.0, a_1 = 1.0, a_2 = 0.2, \beta = 0.6, v_\mu = v_\eta = 0.04, \Omega = 0.4; \delta_\mu = 0.5$ for ADN and $p_\eta = 0.2$ for STN.

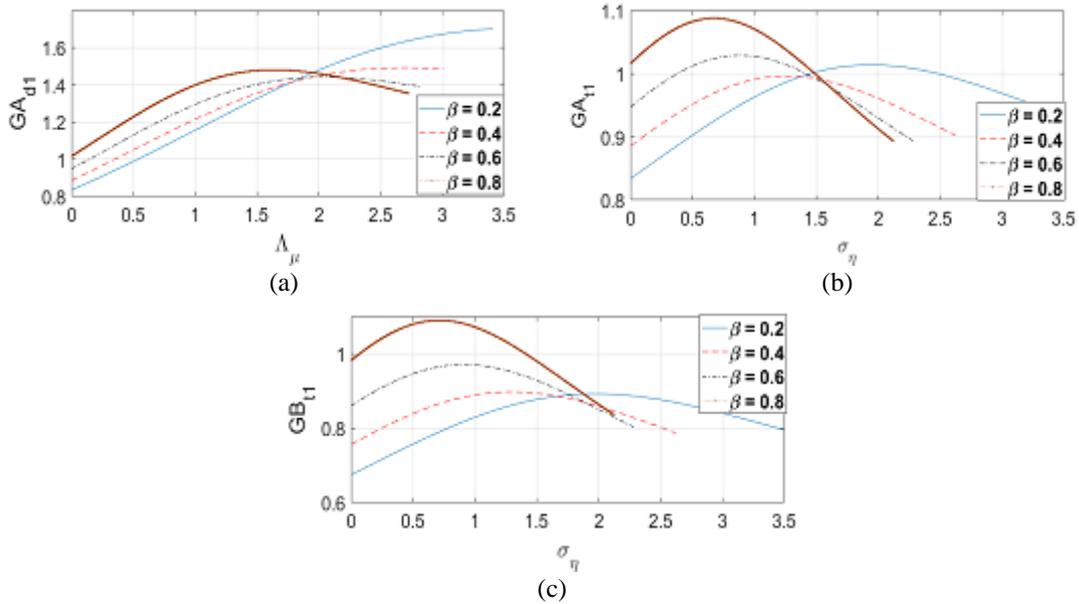

**Fig. 2.** Effect of variation in $\beta$ values on different gains vs noise intensity plots: (a) $GA_{d1}$, (b) $GA_{t1}$, and (c) $GB_{t1}$. The values used for the remaining parameters are $m = 1.0, \gamma = 0.4, k = 1.0, a_1 = 1.0, a_2 = 0.2, \alpha = 1.4, v_\mu = v_\eta = 0.04, \Omega = 0.4; \delta_\mu = 0.5$ for ADN and $p_\eta = 0.2$ for STN.

Furthermore, if these noises are not correlated, then the right-hand side of expressions for $GB_{d1}$ and $GB_{t1}$ in Eqs. (31) and (58), respectively, will be zero implying no effect of the modulated applied force on the system. Consequently, the noises $\xi(t)$ and $\zeta(t)$ should be correlated with $\mu(t)$ and $\eta(t)$, respectively, but not be the same.

In order to find out the influence of the order of fractional derivatives on the gains as function of noise intensity, we have studied these for different non-integer values of $\alpha$ (taking $\beta = 0.6$) and of $\beta$ (keeping $\alpha = 1.4$), and have projected typical graphs in Figs. 1 and 2, respectively. The values used for other parameters are listed in the figure captions. In the case of ADN, the plots for the modulated applied force have not been included in both the figures because the basic information contained is the same as that in Figs. 1(a) and 2(a) for $GA_{d1}$. In fact, the peak values of $GB_{d1}$ vs $\Lambda_\mu$ plots are nearly 54 – 57 % of the corresponding $GA_{d1}$ plots for different values of $\alpha$, and 51 – 60 % in the case of variation with $\beta$. In contrast with this, the $GA_{t1}$ and $GB_{t1}$ vs $\sigma_\eta$ plots show reasonably different behaviours as can be seen from the two figures. For lower values of $\alpha$ as well as $\beta$, the peak values of the plots for the modulated applied force are less than the corresponding ones for the non-modulated driving force, and the trend changes for higher values of both the orders of fractional derivative.

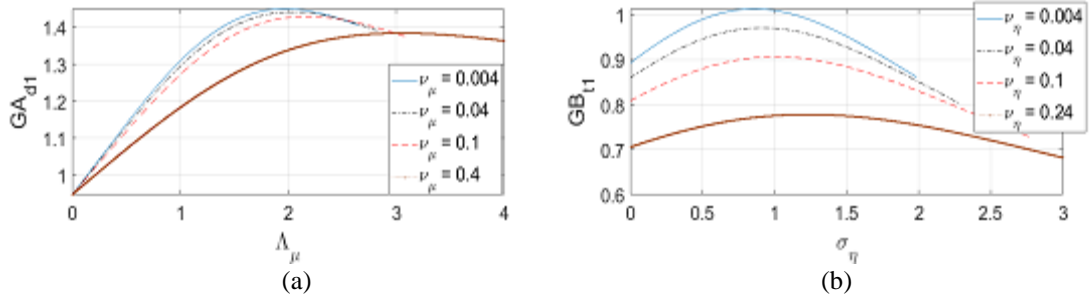

(a) (b)

**Fig. 3.** Dependence of Gains vs noise intensity for (a) $GA_{d1}$ on $v_\mu$, and (b) $GB_{t1}$ on $v_\eta$. Other parameter values used are $m = 1.0, \gamma = 0.4, k = 1.0, a_1 = 1.0, a_2 = 0.2, \alpha = 1.4, \beta = 0.6, \Omega = 0.4; \delta_\mu = 0.5$ for ADN and $p_\eta = 0.2$ for STN. $(\Lambda_\mu)_{cr}$ for $v_\mu = 0.4$ has been taken to be 4.0 rather than the actual value 5.36 and $(\sigma_\eta)_{cr}$ for $v_\eta = 0.24$ as 3.0 instead of actual value 4.88 so that the remaining three plots appear to be well separated.

It is well known, and is also clear from Figs. 1 and 2, that lower values of $\alpha$ and $\beta$ significantly reduce the gains. Also, it has been found that when both $\alpha$ and $\beta$ are taken close to their respective lower limits, such as $\alpha = 1.2, \beta = 0.2$, the $GA_{d1}$ vs $\Lambda_\mu$ plot exhibits only monotonous behaviour though the other graphs do show SR peaks. Guided by these observations and to properly highlight the memory effects, which vary inversely with the derivative-order, we have used $\alpha = 1.4$ and $\beta = 0.6$ for both ADN and STN, in the numerical calculations to follow.

Next, we have assessed the effect of correlation rate of the two noises on the gains vs noise intensity plots by studying these for different values of $v_\mu$ and $v_\eta$. It has been found that for $m = 1.0, \gamma = 0.4, k = 1.0, a_1 = 1.0, a_2 = 0.2, \alpha = 1.4, \beta = 0.6, \delta_\mu = 0.5, (\Lambda_\mu)_{cr}$ is a complex number for $v_\mu \geq 0.455$ for ADN, while for STN with $p_\eta = 0.2$ and other parameters as above, $(\sigma_\eta)_{cr}$ is complex when $v_\eta \geq 0.259$. In this case, the plots for $GA_{d1}$ vs $\Lambda_\mu$ for a chosen set of $v_\mu$ values and for $GB_{t1}$ vs $\sigma_\eta$ have been displayed in Fig. 3. The $GB_{d1}$ vs $\Lambda_\mu$ plots have been excluded because these show similar behaviour as for $GA_{d1}$ in Fig. 3(a) except that the peak values are lower, which decrease from about 57 % for $v_\mu = 0.004$ to nearly 43 % for $v_\mu = 0.4$. Furthermore, the $GA_{t1}$ vs $\sigma_\eta$ plots have been omitted as for the allowed values of $v_\eta$ these are not much different from each other. In fact, the peak values of $GA_{t1}$ for $v_\eta = 0.004$. 0.04, and 0.24 are 1.026, 1.029, and 1.046, respectively; obviously, these are not much sensitive to change in the $v_\eta$ values. Keeping in mind all these aspects and the fact that the noise correlation time is neither very small nor very large, we have chosen $v_\mu = v_\eta = 0.04$ for the calculations in this work.

Now on wards, we shall essentially concentrate on discussing the outcome of our investigation of the influence of variation in the mass, the friction and the potential parameters on both types of gains for the system driven by ADN and STN. However, it is important to note that numerators and denominators of the expressions for the squares of the gains contain polynomials of different orders in $m, \gamma$, and $k$ so that it is not possible to anticipate the exact nature of dependence in an analytical form. Nonetheless, the order of polynomials in the denominator is higher in all the cases implying that the plots for a chosen gain will be lower for the respective higher value of $m, \gamma$, and $k$. In order to concretize this aspect, we have determined the peak value of different gains for various values of $m, \gamma$, and $k$, by varying one of these and keeping the other two fixed, and have obtained the best fit of the results to $m^{-e_m}, \gamma^{-e_\gamma}$ and $k^{-e_k}$, respectively, for different combination of other parameters, to find the corresponding exponent values.

Before proceeding further, it may be mentioned that the plots for various gains shown in the sequel are based on typical combination of the parameter values $a_2 = 0.0, 0.2; \delta_\mu = 0.0, 0.5; p_\eta = 0.5, 0.2$ for $\alpha = 1.4, \beta = 0.6, a_1 = 1.0, v_\mu = v_\eta = 0.04, \Omega = 0.4$; and other related values listed at appropriate places. The choice of two different values of $\delta_\mu$ brings out the effect of asymmetry of ADN, while that of $p_\eta$ values tells us about the effect of probability factor in the case of STN. Similarly, findings for nonzero value of $a_2$ enable us to learn about the relevance of quadratic term in the noise.

### 5.1 Effect of variation in $m$

For $\gamma = 0.4, k = 1.0$ and other parameter values listed in the preceding paragraph, most of the plots for $GA_{d1}, GB_{d1}$ and $GA_{t1}, GB_{t1}$ as function of $\Lambda_\mu$ or $\sigma_\eta$ show non-monotonous behavior only when $m$ lies between 0.6 and 2.4; for some choice of other parameter values the higher value of $m$ is even less than 1.8. Moreover, when $\gamma$ is taken to be 1.0, the lower limit of $m$ becomes 0.8 and upper limit is also reduced. It is worth noting that the expressions for $(\Lambda_\mu)_{cr}$ and $(\sigma_\eta)_{cr}$, involve $mv_\mu^\alpha$ and $mv_\eta^\alpha$, respectively, which, for the values used here, become $0.011\,m$ — a factor much smaller than the sum of the other factors mainly because of the presence of $k = 1.0$. Consequently, for a specific choice of other parameter values, both $(\Lambda_\mu)_{cr}$ and $(\sigma_\eta)_{cr}$ do not exhibit significant variation with change in $m$.

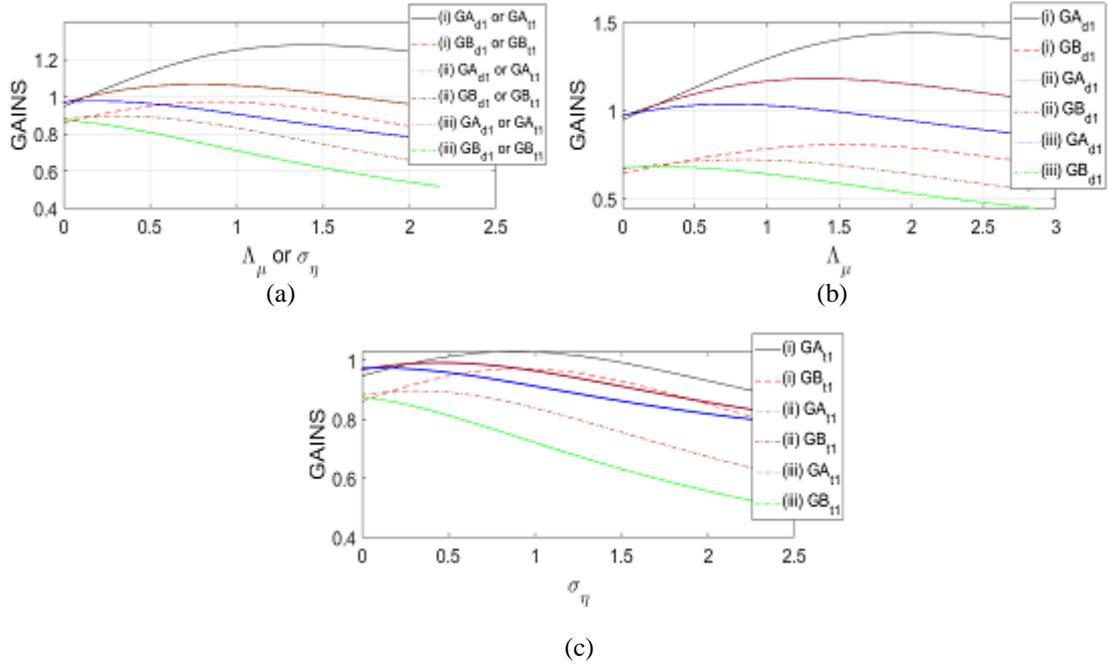

**Fig. 4.** Plots of gains vs noise intensities for (i) $m = 1.0$, (ii) $m = 1.5$, (iii) $m = 2.0$; and $\gamma = 0.4$, $k = 1.0$, $a_1 = 1.0$, $a_2 = 0.2$, $\alpha = 1.4$, $\beta = 0.6$, $v_\mu = v_\eta = 0.04$, $\Omega = 0.4$; corresponding to (a) $\delta_\mu = 0.0$ or $p_\eta = 0.5$; (b) $\delta_\mu = 0.5$ ; (c) $p_\eta = 0.2$.

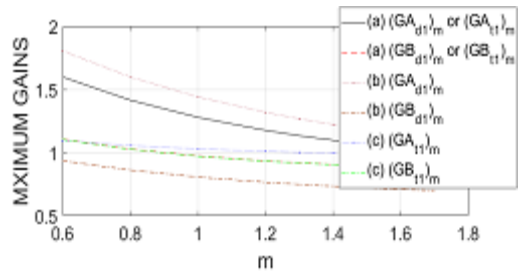

**Fig. 5.** Dependence of maxima in the Gains vs noise intensities on $m$ for other parameter values as listed in Fig.4.

Fig. 4 shows the plots for different gains corresponding to three values of mass parameter, viz. $m = 1.0$, 1.5 and 2.0, and (a) $\delta_\mu = 0.0$ or $p_\eta = 0.5$, (b) $\delta_\mu = 0.5$, (c) $p_\eta = 0.2$, and other typically chosen parameters listed in the caption there. A perusal of these figures reveals that SR for a chosen set of parameter values becomes less prominent with increase in $m$ and this trend is the least for $\delta_\mu = 0.5$ and the highest for $p_\eta = 0.2$. Also, the magnitudes of the peak values are maximum for $\delta_\mu = 0.5$ and minimum for $p_\eta = 0.2$ with the results for $\delta_\mu = 0.0$ or $p_\eta = 0.5$ in-between. In fact, the SR peak positions shift towards lower noise intensity and peak magnitudes decrease with increase in $m$ and these changes are influenced by the values of $a_2$, $\delta_\mu$ and $p_\eta$.

The decrease in the peak value with increase in $m$ has been projected in figure 5. The $GB_{t1}$- peak curves for $p_\eta = 0.2$ are very close to those for $p_\eta = 0.5$ whether $a_2 = 0$ or 0.2, indicating that the $m$ dependence of gains corresponding to the modulated applied force is not much sensitive to the value of $p_\eta$. It may be mentioned that, in general, the peak values of all the gains for $a_2 = 0.0$ are higher than those for $a_2 = 0.2$ when the other parameters are the same, and the difference in these is more pronounced in the case of $GB_{d1}$ and $GB_{t1}$ as compared to $GA_{d1}$ and $GA_{t1}$ for which the difference is relatively small.

As mentioned earlier, we have specified the trend of variation in the peak value of different gains with $m$, by obtaining the best fit of the results to $m^{-e_m}$ for various combinations of other parameters and the exponents $e_m$ so obtained have been listed in Table 1. A look at the entries in this table brings out the fact that for the non-modulated applied force, $(e_m)_A$ is somewhat smaller for $a_2 = 0.0$ than that for $a_2 = 0.2$ for ADN, SDN, or STN with $p_\eta = 0.5$. However, for STN with $p_\eta = 0.2$, the values of $(e_m)_A$ are quite close to each other for $a_2 = 0.0$ and $a_2 = 0.2$, and their values are much smaller than those for ADN, SDN, or STN with $p_\eta = 0.5$. On the other hand, in the case of the results for noise-modulated driving force, the exponents for $a_2 = 0.0$ are much larger than those for $a_2 = 0.2$, whether $\delta_\mu = 0.0$ or $0.50$ and $p_\eta = 0.2$ or $0.5$; in fact, for STN the values of $(e_m)_B$ are independent of the value of the probability parameter as anticipated on the basis of the plots in Fig. 5.

**Table 1.** Exponents for variation of the peak values of different gains with $m$, $\gamma$ and $k$ for $a_1$ = 1.0, $\alpha$ = 1.4, $\beta$ = 0.6, $v_\mu = v_\eta = 0.04$, $\Omega = 0.4$; other parameter values are : (i) $a_2 = 0.0$, $\delta_\mu = 0.0$ or $p_\eta = 0.5$; (ii) $a_2 = 0.0$, $\delta_\mu = 0.5$; (iii) $a_2 = 0.2$, $\delta_\mu = 0.0$ or $p_\eta = 0.5$; (iv) $a_2 = 0.20$, $\delta_\mu = 0.50$; (v) $a_2 = 0.0$, $p_\eta = 0.2$; (vi) $a_2 = 0.2$, $p_\eta = 0.2$.

| Identifying Number | $(e_m)_A$ $\gamma = 0.4$, $k = 1.0$ | $(e_m)_B$ $\gamma = 0.4$, $k = 1.0$ | $(e_m)_A$ $\gamma = 1.0$, $k = 1.0$ | $(e_m)_B$ $\gamma = 1.0$, $k = 1.0$ | $(e_\gamma)_A$ $m = 1.0$, $k = 1.0$ | $(e_\gamma)_B$ $m = 1.0$, $k = 1.0$ | $(e_k)_A$ $m = 1.0$, $\gamma = 0.4$ | $(e_k)_B$ $m = 1.0$, $\gamma = 0.4$ |
|---|---|---|---|---|---|---|---|---|
| (i) | 0.41 | 0.41 | 0.20 | 0.23 | 0.52 | 0.83 | 0.22 | 1.01 |
| (ii) | 0.48 | 0.45 | 0.27 | 0.27 | 0.56 | 0.60 | 0.22 | 0.78 |
| (iii) | 0.43 | 0.23 | 0.18 | 0.028 | 0.54 | 0.59 | 0.22 | 1.38 |
| (iv) | 0.47 | 0.27 | 0.23 | 0.050 | 0.59 | 0.60 | 0.26 | 1.12 |
| (v) | 0.11 | 0.40 | 0.056 | 0.24 | 0.38 | 0.83 | 0.63 | 1.01 |
| (vi) | 0.10 | 0.23 | 0.035 | 0.028 | 0.36 | 0.61 | 0.65 | 1.34 |

When $\gamma$ is increased from 0.4 to 1.0 leading to higher frictional or external damping; as expected, not only the range of $m$ values yielding non-monotonous behaviour of the gains is curtailed, but also the maximum values of the gains are reduced. The exponents for this case have also been included in Table 1. It is found that for the same set of other parameter values, the decrease in the peak magnitudes is relatively less as compared to the smaller friction parameter.

*5.2 Effect of variation in $\gamma$*

With a view to bring out the effect of change in the friction parameter values on different gains, we have depicted plots for gains corresponding to $\gamma = 0.8$ and 1.2 in Fig. 6 choosing other parameter values as mentioned there. A scrutiny of these plots together with the relevant ones, indicated as (i) in Fig. 4 for $\gamma = 0.4$, shows that rise in the $\gamma$ value decreases the gain because the fractional external damping increases at the cost of the fractional internal damping. Here, too, the influence is also dependent on the values of $a_2$, $\delta_\mu$ and $p_\eta$.

It may be mentioned that except when the system is under the influence of STN ($p_\eta = 0.2$), the peaks in the gains obtained with the non-modulated applied force have higher values than those for the corresponding modulated force driven cases irrespective of whether quadratic term in the noise is present or not. In contrast, for STN ($p_\eta = 0.2$, $a_2 = 0.0$ or 0.2), $GB_{t1} > GA_{t1}$ for small values of $\gamma$ and the relative magnitudes get interchanged for larger $\gamma$. This happens for $\gamma = 0.593$ when $a_2 = 0.0$ and for $\gamma = 0.325$ if $a_2 = 0.2$.

The exponents $e_\gamma$ accounting for the trend of variation of the peak values in the plots for various gains versus relevant noise intensity, with $\gamma$ have been also shown in Table 1. We have used $\gamma = 0.4$ as the lowest value for determining these exponents because inclusion of SR peak values for $\gamma < 0.4$ drastically affects the quality of the fit. It is found that $(e_\gamma)_A$ have reasonably consistent value of about 0.55 for ADN ($\delta_\mu = 0.0$ or 0.5) as well as STN ($p_\eta = 0.5$), and 0.37 for STN ($p_\eta = 0.2$), whether $a_2 = 0.0$ or 0.2. However, in the case of modulated applied force, the exponent $(e_\gamma)_B$ has value 0.83 for SDN, STN ($p_\eta = 0.5, 0.2$) for $a_2 = 0.0$, while its value for all other cases is nearly 0.60.

Next, when we take $\gamma$ to be quite small, 0.01, so that the fractional internal damping term is much larger than the external damping term, the behaviour of the system becomes significantly different. The peak values corresponding to $\Omega = 0.4$ are at substantial variance with the values obtained using the relevant exponents listed in Table 1. Furthermore, the gains become high as $\Omega$ is reduced and the noise intensity for which the SR peak is observed, moves closer to the corresponding critical value. As a typical example, we have depicted in Fig. 7, the graphs for $\Omega = 0.01$ for the system subjected to STN ( $p_\eta = 0.2$), for which $(\sigma_\eta)_{cr} = 1.9726$ and the peaks occur for $\sigma_\eta = 1.907$. We have shown these graphs for a limited range of $\sigma_\eta$ values to highlight this aspect. It may be added here that if we take $a_2 = 0.0$, keeping other parameters the same as in Fig. 7, the peaks of reasonably higher magnitude are obtained at $\sigma_\eta = 1.0194$, while $(\sigma_\eta)_{cr} = 1.0200$.

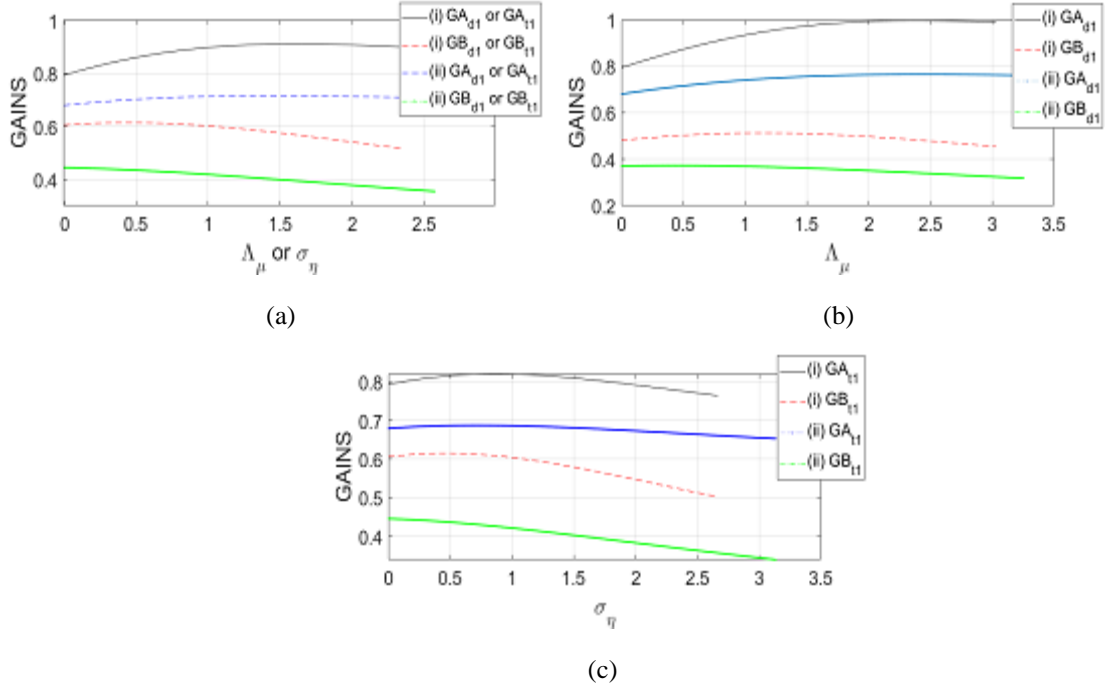

**Fig. 6.** Gains as function of relevant noise intensities for (i) $\gamma = 0.8$, (ii) $\gamma = 1.2$ for $m = 1.0$, $k = 1.0$, $a_1 = 1.0$, $a_2 = 0.2$, $\alpha = 1.4$, $\beta = 0.6$, $v_\mu = v_\eta = 0.04$, $\Omega = 0.4$; (a) $\delta_\mu = 0.0$ or $p_\eta = 0.5$; (b) $\delta_\mu = 0.5$ ; (c) $p_\eta = 0.2$.

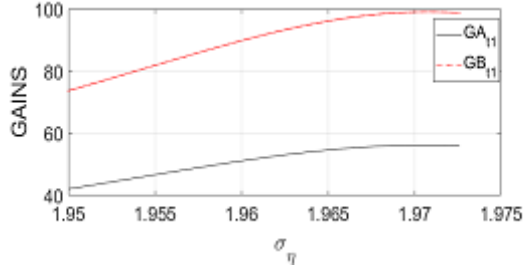

**Fig. 7.** Plot of Gains vs noise intensity under STN for $\gamma = 0.01$; other parameter values are $m = 1.0$, $k = 1.0$, $a_1 = 1.0$, $a_2 = 0.2$, $\alpha = 1.4$, $\beta = 0.6$, $v_\eta = 0.04$, $\Omega = 0.01$, $p_\eta = 0.2$.

*5.3 Effect of variation in potential parameter*

In order to analyse the effect of variation in the potential parameter $k$ on the gains, we have taken $m = 1.0$ and $\gamma = 0.4$. It is found that for the parameter values commonly used so far and constituting the content of caption for Fig. 8, the gains exhibit non-monotonous behaviour if $k$ lies between 0.6 and 1.2. Also, some of the critical values of noise intensity turn out to be complex numbers for $k > 1.2$. The plots for gain vs noise intensity pertaining to $k = 0.6$ and 0.8 are shown in Fig. 8. These together with the entries (i) in Fig. 4, reveal that the SR peak magnitudes decrease with increase in the $k$ values. It is pertinent to note that except the case when $\delta_\mu = 0.5$, the peak values for the system under the influence of the modulated external force are higher than those for the non-modulated driving force irrespective of the value of $a_2$, for lower values of $k$. However, this situation changes at higher values of $k$; the value for which this situation gets changed, is different for different sets of parameter values. For example, when $a_2 = 0.2$, $\delta_\mu = 0.0$ or $p_\eta = 0.5$, this happens for $k = 0.790$, while for $p_\eta = 0.2$ the corresponding $k$ value is found to be 0.929. Interestingly, for $a_2 = 0.0$, $\delta_\mu = 0.0$ or $p_\eta = 0.5$ the change takes place for $k = 0.889$ and the gain vs noise intensity plots completely overlap each other implying exactly same behaviour of the gains

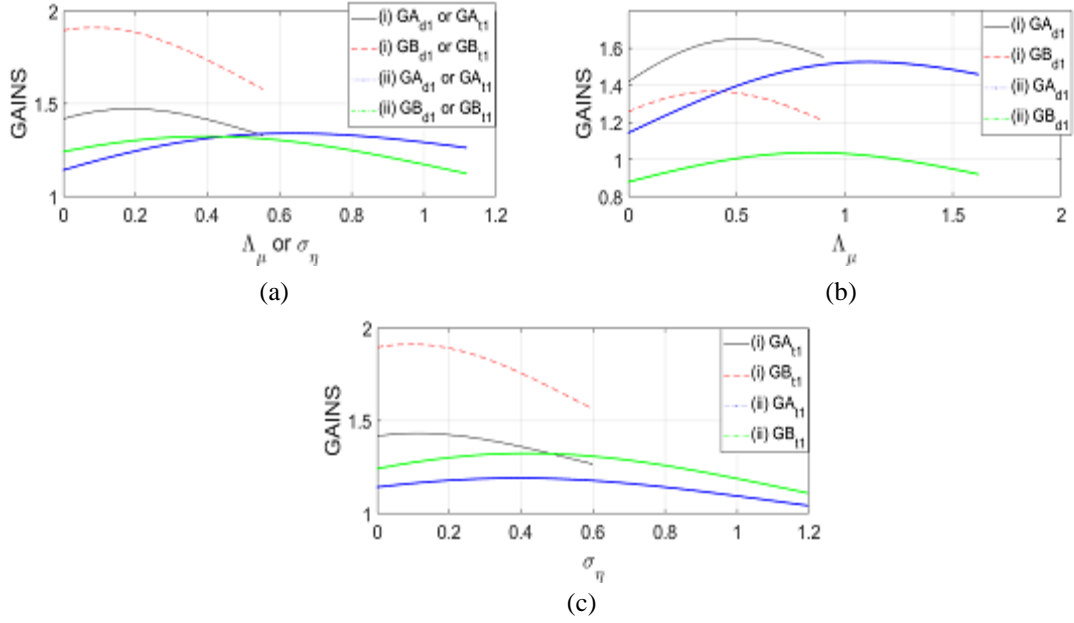

**Fig. 8.** Graphs of gains vs noise intensities for (i) $k = 0.6$, and (ii) $k = 0.8$; values of other parameters used are $m = 1.0, \gamma = 0.4, a_1 = 1.0, a_2 = 0.2, \alpha = 1.4, \beta = 0.6, \upsilon_\mu = \upsilon_\eta = 0.04, \Omega = 0.4$; (a) $\delta_\mu = 0.0$ or $p_\eta = 0.5$; (b) $\delta_\mu = 0.5$; (c) $p_\eta = 0.2$.

in the absence of the quadratic term in the presence of SDN or STN ($p_\eta = 0.5$).

The exponents $e_k$ that describe the dependence of SR peaks on $k$ too have been included in Table 1. In this case also, the $(e_k)_A$ values are fairly harmonious with each other for ADN ($\delta_\mu = 0.0$ or $0.5$) and similarly for STN ($p_\eta = 0.2$) for both $a_2 = 0.0$ and $0.2$. But for $(e_k)_B$, such a general statement does not hold good; all we can say is that these have higher values in the presence of the quadratic noise.

As a follow up of the inference drawn in the paragraph after Eq. (36), we have examined the behaviour of the system driven by quadratic ADN for $k = 0$ and $a_2 = 0.2$. For the parameter values generally used here, $(\Lambda_\mu)_{cr}$ has been found to be 24.655 and 27.155 for $\delta_\mu = 0.0$ and $0.5$, respectively. As argued earlier, the SR peaks are obtained only for lower values of $\Omega$ and the results for $\Omega = 0.1, 0.01$ and $0.001$ are depicted in Fig. 9. The result for $\Omega = 0.1, \delta_\mu = 0.0$ has not been included in Fig. 9(a) because the plot for $GB_{d1}$ shows monotonous behaviour and that for $GA_{d1}$, it is not much different from that for $\delta_\mu = 0.5$. It may be noted that for a particular $\Omega$, the SR peaks in $GA_{d1}$ and $GB_{d1}$ occur for nearly the same value of $\Lambda_\mu$ and that these values increase as $\Omega$ decreases; these become close to the corresponding $(\Lambda_\mu)_{cr}$. Also, the separation between the graphs for $GA_{d1}$ and $GB_{d1}$ becomes less with decrease in the $\Omega$ values and, hence, the $\Lambda_\mu$ scale for Fig. 9(b) and 9(c) has been taken to be different from that in Fig. 9(a). It is worth mentioning that the values of the two gains get enhanced by a factor of about 5 when $\Omega$ is decreased by a factor of 10. Also, all the plots exhibit broad valleys, which is distinctive feature of the $k = 0$ case. As mentioned earlier, such a situation is not possible when the system is under the influence of STN.

## 6. Numerical simulations

With a view to examine the reliability of the analytical expressions obtained for the collective SR and accuracy of the results so obtained, we have numerically solved Eq. (4), omitting the term containing drift force as it makes no contribution to SR, for some combinations of different parameters and have determined values for the gains at the relevant peak positions through Monte Carlo experiment.

In order to achieve our goal, we rewrite Eq. (4) as

$$mD^\alpha X(t) + \gamma D^\beta X(t) = \mathcal{F}(X(t), t), \tag{67}$$

where
$$\mathcal{F}(X(t), t) = -[k + \{a_1 + a_2 \delta_\mu\}\mu(t) + a_2 \Lambda_\mu]X(t) + (1 + c)[A + B\xi(t)]\sin(\Omega t) \tag{68a}$$

for the ADN, and

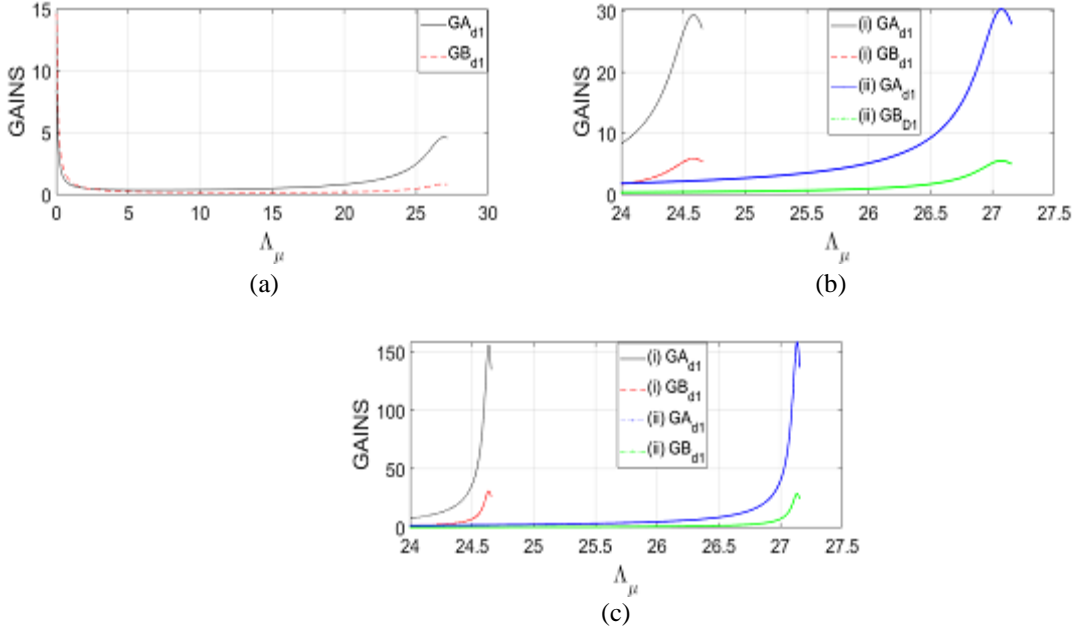

**Fig. 9.** Dependence of gains vs noise intensity on $\Omega$ for $k = 0.0$ in the case of ADN with other parameters as $m = 1.0, \gamma = 0.4, a_1 = 1.0, a_2 = 0.2, \alpha = 1.4, \beta = 0.6, v_\mu = v_\eta = 0.04$; (a) $\Omega = 0.1$, $\delta_\mu = 0.5$; (b) $\Omega = 0.01$ and (i) $\delta_\mu = 0.0$, (ii) $\delta_\mu = 0.5$; (c) $\Omega = 0.001$, (i) $\delta_\mu = 0.0$, (ii) $\delta_\mu = 0.5$.

$$\mathcal{F}(X(t), t) = -[k + a_1\eta(t) + a_2\sigma_\eta]X(t) + (1 + c)[A + B\zeta(t)]\sin(\Omega t) \qquad (68b)$$

in the case of STN. Note that Eq. (6) has been used while writing Eq. (68a).

Guided by the finite-difference method based time-discretization scheme for the Caputo form of the fractional derivative developed by Lin and Xu [53] for $0 < \beta \leq 1$, we introduce discrete time $t_l = l\Delta t$, where $\Delta t$ is the time–step size. Thus, the discretized form of Eq. (67) reads

$$mD^\alpha X(t_{L+1}) + \gamma D^\beta X(t_{L+1}) = \mathcal{F}(X(t_{L+1}), t_{L+1}), \qquad (69)$$

with

$$mD^\alpha X(t_{L+1}) = \left[\frac{m}{\Gamma(2-\alpha)}\right]\sum_{l=0}^{L}\frac{X(t_{l+1}) - 2X(t_l) + X(t_{l-1})}{(\Delta t)^2}\int_{t_l}^{t_{l+1}}(t_{L+1} - t')^{1-\alpha}dt' \qquad (70)$$

and

$$\gamma D^\beta X(t_{L+1}) = \left[\frac{\gamma}{\Gamma(1-\beta)}\right]\sum_{l=0}^{L}\frac{X(t_{l+1}) - X(t_l)}{\Delta t}\int_{t_l}^{t_{l+1}}(t_{L+1} - t')^{-\beta}dt'. \qquad (71)$$

Here, we have ignored the local truncation error which can be made negligible by choosing $\Delta t$ to be sufficiently small.

Evaluating the integrals in the preceding two equations, rearranging various terms in these, and substituting the resulting expressions into Eq. (69), where $\mathcal{F}(X(t_{L+1}), t_{L+1})$ is replaced by $\mathcal{F}(X(t_L), t_L)$, we finally get general expression for the numerical solution of Eq. (4). We have then used the algorithms prescribed by Barik et al [54], and Zhou and Lin [55] to generate the ADN and STN, respectively, needed for determining the value of the gains for different combinations of the parameters.

To begin with, the problem was considered in the absence of noise for $m = k = 1.0, \gamma = 0.4, \alpha = 1.4, \beta = 0.6$, and $\Omega = 0.4$. The analytic expressions, Eqs. (29a) and (56a), gave the gain $GA_{an} = 0.9470$, while the numerical solution for $T = 60$ led to $GA_{num} = 0.9474$ (difference in %, $\Delta = 0.042$) and $0.9472$ ($\Delta = 0.021$) for $\Delta t = 0.01$ and $0.0075$, respectively; however, for $T = 20$ and $\Delta t = 0.0075$, $GA_{num}$ turned out to be $0.9499$ ($\Delta = 0.31$). Guided by these observations, all the numerical solutions have been obtained using $T = 60$ and time–step size $\Delta t = 0.0075$. Furthermore, the Monte Carlo experiment has been executed with 11000 realizations of the noise for each case. The numerical results have been compared with the corresponding analytical values of the gains for the peak value of the gain vs noise intensity plots, in Table 2. The good quality of agreement between the two categories of values

**Table 2.** A comparison of the peak–value analytical and numerical simulation results for the gains for different combinations of $a_2, \delta_\mu$ or $p_\eta$ values for $m$ = 1.0, $\gamma$ = 0.4, $k$ = 1.0, $a_1$ = 1.0, $a_2$ = 0.2, $\alpha$ = 1.4, $\beta$ = 0.6, $v_\mu = v_\eta$ = 0.04, $\Omega$ = 0.4.

| $a_2, \delta_\mu, p_\eta$ values | Type of gain | Analytical result | Numerical result | % Difference ($\Delta$) |
|---|---|---|---|---|
| $a_2 = 0.0$, $\delta_\mu = 0.0$ | $GA_{d1}$ | 1.325 | 1.316 | 0.68 |
| | $GB_{d1}$ | 1.203 | 1.190 | 1.08 |
| $a_2 = 0.0$, $\delta_\mu = 0.5$ | $GA_{d1}$ | 1.518 | 1.507 | 0.72 |
| | $GB_{d1}$ | 0.964 | 0.978 | 1.45 |
| $a_2 = 0.2$, $\delta_\mu = 0.0$ | $GA_{d1}$ | 1.279 | 1.268 | 0.86 |
| | $GB_{d1}$ | 0.973 | 0.968 | 0.51 |
| $a_2 = 0.2$, $\delta_\mu = 0.5$ | $GA_{d1}$ | 1.441 | 1.434 | 0.49 |
| | $GB_{d1}$ | 0.807 | 0.814 | 0.87 |
| $a_2 = 0.0$, $p_\eta = 0.5$ | $GA_{t1}$ | 1.325 | 1.310 | 1.13 |
| | $GB_{t1}$ | 1.203 | 1.193 | 0.83 |
| $a_2 = 0.0$, $p_\eta = 0.2$ | $GA_{t1}$ | 1.048 | 1.058 | 0.95 |
| | $GB_{t1}$ | 1.205 | 1.187 | 1.49 |
| $a_2 = 0.2$, $p_\eta = 0.5$ | $GA_{t1}$ | 1.279 | 1.270 | 0.70 |
| | $GB_{t1}$ | 0.973 | 0.964 | 0.92 |
| $a_2 = 0.2$, $p_\eta = 0.2$ | $GA_{t1}$ | 1.029 | 1.018 | 1.07 |
| | $GB_{t1}$ | 0.971 | 0.988 | 1.75 |

listed here shows that the analytical findings are completely trustworthy.

## 7. Concluding remarks

This work is devoted to detailed investigation of SR in a system consisting of a finite number of globally coupled oscillators with double fractional-order damping, in the presence of multiplicative quadratic asymmetric dichotomous or symmetric trichotomous noise affecting the potential parameter, the coupling factor and the local drift force (the noise is same in these cases but differs in extent of influence on these parameters) and driven by a sinusoidal force which is either noise-free or noise-modulated. The effect of interaction between the applied field and the heat bath has been incorporated through a simple model presented in the appendix. The problem has been analysed in terms of mean field so that the study pertains to collective response of the system. The fractional stochastic equation describing the instantaneous mean displacement under the influence of specific noise has been transformed into a set of deterministic differential equations, which have been solved using the Laplace transform method. The exact analytical expressions so obtained, have been first used to bring out the effect of the order of fractional derivatives as well as the correlation rates and then to systematically look into the influence of variation in mass, friction, and potential parameters on the collective system output amplitude gains as function of noise-intensity (less than the relevant critical noise intensity in conformity with the relevant stability condition). These investigations have been carried out for two values each of asymmetry parameter for ADN and probability parameter for STN. The main conclusions drawn are as following.

i) The noise modulating the applied periodic force should be correlated with the noise affecting other parameters but should not be the same.
ii) When mass parameter of the oscillators is increased, the value of noise intensity for which SR peak is obtained, decreases and so does the value of maximum gain. Both these and the exponent governing the variation of peak value with $m$ depend on the values of $a_2$, $\delta_\mu$ and $p_\eta$, the nature of the noise and, also, whether the external force is non-modulated or modulated by noise. These aspects have been explored for two different values of friction parameter.
iii) The increase in the value of friction parameter does reduce the SR peak value but it does not follow a specific trend for the variation in peak position. Once again, these variations and the exponent describing the dependence of peak magnitude on the friction parameter depend on other parameters and whether the noise is ADN or STN. For sufficiently low value of $\gamma$, the SR peak magnitudes significantly differ from the values found with the relevant exponents. Also, high magnitude peaks are obtained when the applied force frequency is decreased by an order of magnitude, but their location turns out to be quite close to the corresponding critical noise intensity.

iv) Besides analysing various aspects of the effect of variation in potential parameter on the collective SR, we have also determined the exponents accounting for the dependence of SR peaks on this. Their values for the non-modulated applied force are quite consistent with each other for both ADN and STN irrespective of the values of other parameters. However, in the case of modulated force this is not so.

v) It has been argued that the presence of the second-order term in the multiplicative ADN associated with $X(t)$ can produce oscillatory motion even when $k = 0$. Such a system has large value of critical noise intensity and, as anticipated, shows SR at lower values of the frequency of the driving sinusoidal force; the plots also have broad valleys. However, a system subjected to STN cannot exhibit this feature.

vi) A comparison of the numerical simulations-based results pertaining to different gain peaks for some typical cases with the corresponding findings of the analytical calculations brings out the efficacy of the latter.

## Appendix A. Expression for the effect of external time-dependent field on the harmonic oscillator and the heat bath

We consider a harmonic oscillator having mass $m_0$, potential parameter $k_0$ defining its intrinsic frequency, instantaneous position $x_0(t)$ and instantaneous momentum $p_0(t)$; and coupled with all the $N_b$ constituents of the independent-oscillators-heat-bath, characterized by masses $\{m_j\}$, instantaneous positions and momenta $\{q_j(t)\}$ and $\{p_j(t)\}$, respectively, with $j = 1,2,\ldots,N_b$. We assume that the coupling between the central oscillator and the bath oscillators is through linear springs of force constant $\{k_j = m_j \omega_j^2\}$ and that a time-dependent external driving force $F(t)$ interacts with both the tagged oscillator and the bath oscillators. Thus, the Hamiltonian for the composite driven system reads [41,42]

$$H = \frac{p_0^2}{2m_0} + \frac{k_0 x_0^2}{2} + \sum_{j=1}^{N_b}\left\{\frac{p_j^2}{2m_j} + \frac{m_j \omega_j^2}{2}(x_0 - q_j)^2\right\} - F(t)\left\{x_0 + \sum_{j=1}^{N_b} c_j q_j\right\}. \tag{A1}$$

We have taken the force $F(t)$ to be such that its coupling parameter with $x_0(t)$ is unity and the corresponding coefficients with $\{q_j(t)\}$ are $\{c_j\}$.

Following the usual method (see, e.g., [39]), we derive Hamilton's canonical equations for the tagged oscillator as well as the $N_b$ bath oscillators, carry out an integration by parts in the latter, and finally obtain the following generalized Langevin equation to describe the motion of the central oscillator :

$$m_0 \frac{d^2 x_0(t)}{dt^2} + \int_0^t \gamma(t - t') \frac{dx_0(t')}{dt'} dt' + k x_0(t) = \Gamma(t) + F_{\text{eff}}(t). \tag{A2}$$

Here,

$$\gamma(t) = \sum_{j=1}^{N_b} m_j \omega_j^2 \cos(\omega_j t), \tag{A3}$$

$$\Gamma(t) = \sum_{j=1}^{N} m_j \omega_j^2 \left[\{q_j(0) - x_0(0)\}\cos(\omega_j t) + \frac{dq_j(0)}{dt}\frac{\sin(\omega_j t)}{\omega_j}\right], \tag{A4}$$

and

$$F_{\text{eff}}(t) = F(t) + F_b(t) \tag{A5}$$

with

$$F_b(t) = \sum_{j=1}^{N_b} c_j \omega_j \int_0^t F(t') \sin[\omega_j(t - t')] dt'. \tag{A6}$$

Note that $\gamma(t)$ is the damping kernel for time-retarded friction, $\Gamma(t)$ is the stochastic force which acts as additive noise, and $F_{eff}(t)$ is the effective driving force acting on the tagged oscillator and consists of two parts: (i) the directly coupled actual applied force $F(t)$ and (ii) the time–retarded force $F_b(t)$ with kernel $\sum_{j=1}^{N_b} c_j \omega_j \sin(\omega_j t)$ arising from the interaction of force $F(t)$ with the bath oscillators and is determined by the coupling coefficients $\{c_j\}$ and the spectral properties of the bath modes $\{\omega_j\}$.

To be specific, we take $F(t)$ to be a sinusoidal force having amplitude $F_0$ and frequency $\Omega$ :

$$F(t) = F_0 \sin(\Omega t). \tag{A7}$$

Accordingly, we have

$$F_b(t) = F_0 \sum_{j=1}^{N_b} c_j \omega_j I_j, \tag{A8}$$

where

$$I_j = \int_0^t \sin(\Omega t') \sin[\omega_j(t-t')] dt'\} = \frac{\Omega \sin(\omega_j t) - \omega_j \sin(\Omega t)}{(\Omega^2 - \omega_j^2)}. \tag{A9}$$

For convenience, we assume that all the bath oscillators are identical in the sense that these have the same frequency and the same coupling to the applied force so that $\omega_j = \omega$ and $c_j = \frac{c}{N_b}$, for all values of $j$. Using this in Eqs. (A8) and (A9), we get

$$F_b(t) = F_0 c \omega \frac{\Omega \sin(\omega t) - \omega \sin(\Omega t)}{(\Omega^2 - \omega^2)}. \tag{A10}$$

It may be pointed out that $F_b(t)$ has $\frac{0}{0}$ indeterminate form for $\omega \to \Omega$, but applying L'Hospital's rule it is found that $F_b(t)$ does not have a singularity for the limit $\omega \to \Omega$. Furthermore, following Cui and Zaccone [39], we assume that $\omega \gg \Omega$, which leads to

$$F_b(t) \approx c F_0 \sin(\Omega t) = c F(t) \tag{A11}$$

and

$$F_{\text{eff}}(t) = (1 + c) F(t). \tag{A12}$$

Obviously, for the model considered here, the coupling of the driving sinusoidal force $F(t)$ to the thermal bath oscillators enhances its effect on the tagged oscillator by a factor equal to the product of coupling parameter and the number of the bath oscillators. Also, $c = 0$ corresponds to the situation that only the central oscillator is experiencing the effect of the applied periodic force.

**Conflict of interest**

The author declares that he has no conflict of interest.

**Funding**

This research did not receive any grant from the funding agencies in the public, commercial, or not-for-profit sectors.

**Acknowledgements**

The author dedicates this work with immense gratitude and respect to Prof S P Puri (on his 90[th] birth anniversary) and to (late) Prof R S Sud, and their families.

**References**

[1] E. Soika, R. Mankin, A. Ainsaar, Resonant behavior of a fractional oscillator with fluctuating frequency. Phys. Rev. E 81 (2010) 011141.
[2] E. Lanzara, R. N. Mantegna, B. Spagnolo, R. Zangara, Experimental study of a nonlinear system in the presence of noise: The stochastic resonance, Amer. J. phys. 65 (1997) 341-349.
[3] L. Gammaitoni, P. Hänggi, P. Jung, F. Marchesoni, Stochastic resonance. Rev. Mod. Phys. 70 (1998) 223-287.
[4] R. N. Mantegna, B. Spagnolo, M. Trapanese, Linear and nonlinear experimental regimes of stochastic resonance, Phys. Rev. E 63 (2001) 011101.
[5] P. Hänggi, Stochastic resonance in biology, Chem. Phys. Chem. 3 (2002) 285-290.
[6] T. Wellens, V. Shatokhin, E. Buchleitner, Stochastic resonance, Rep. Prog. Phys. 67 (2004) 45-105.

[7] R. Benzi, Stochastic resonance: from climate to biology. Nonlinear Processes Geophys. 17 (2010) 431-441.
[8] N. V. Agudov, A. V. Krichigin, D Valenti, B. Spagnolo, Stochastis resonance in a trapping overdamped monostable system, Phys. Rev. E 81 (2010) 051123.
[9] J. C. Li, D. C. Mei, Reverse resonance in stock prices of financial system with periodic information, Phys. Rev. E 88 (2013) 012811.
[10] B. Spagnolo, A. Carollo, D. Valenti, Stabilization by dissipation and stochastic resonant activation in quantum metastable systems, Eur. Phys. J. Sp. Topics 227 (2018) 379-420.
[11] I. Podlubny, Fractional Differential Equations, Academic, San Diego, 1999.
[12] R. Herrmann, Fractional Calculus – An Introduction for Physicists, World Scientific, Singapore, 2014.
[13] A. Tofighi, The intrinsic damping of the fractional oscillator. Physica A 329 (2003) 29-34.
[14] Y. E. Ryabov, A. Puzenko, Damped oscillations in view of the fractional oscillator equation, Phys. Rev. B 66 (2002) 184201.
[15] G. He, Y. Tian, Y. Wang, Stochastic resonance in a fractional oscillator with random damping strength and random spring stiffness, J. Stat. Mech. (2013) P09026.
[16] H. Sun, Y. Zhang, D. Baleanu, W. Chen, Y. Chen, A new collection of real world applications of fractional calculus in science and engineering, Commun. Nonlinear Sci. Numer. Simul. 64 (2018) 213-231.
[17] E. Soika, R. Mankin, Trichotomous-noise-induced stochastic resonance for a fractional oscillator, Proc. WSEAS MABE – 10 (2010) 440-445.
[18] R. Lang, L. Yang, H. Qin, G. Di, Trichotomous noise induced stochastic resonance in a linear system, Nonlinear Dyn. 69 (2012) 1423–1427.
[19] M. Gitterman, The Noisy Oscillator: random mass, frequency, damping. World Scientific, Singapore, 2013.
[20] F. Guo, H. Li, J. Liu, Stochastic resonance in a linear system with random damping parameter driven by trichotomous noise, Physica A 409 (2014) 1-7.
[21] W. Zhang, G. Di, Stochastic resonance in a harmonic oscillator with damping trichotomous noise, Nonlin. Dyn. 77 (2014) 1589-1595.
[22] S. Zhong, H. Ma, H. Peng, L. Zhang, Stochastic resonance in a harmonic oscillator with fractional-order external and intrinsic dampings, Nonlin. Dyn. 82 (2015) 535-545.
[23] L. Lin, C. Chen, H. Wang, Trichotomous noise induced stochastic resonance in a fractional oscillator with random damping and random frequency, J. Stat. Mech. (2016) 023201.
[24] L. Lin, H. Wang, S. Zhong, Stochastic resonance for a fractional oscillator with random trichotomous mass and random trichotomous frequency, Int. J. Mod. Phys. B 31 (2017)1750231.
[25] R. Ren, M. Luo, K. Deng, Stochastic resonance in a fractional oscillator driven by multiplicative quadratic noise, J. Stat. Mech. (2017) 023210.
[26] S. Zhong, L. Zhang, H. Wang, H. Ma, M. Luo, Nonlinear effect of time delay on the generalized stochastic resonance in a fractional oscillator with multiplicative polynomial noise, Nonlinear Dyn. 89 (2017) 1327-1340.
[27] R. Ren, M. Luo, K. Deng, Stochastic resonance in a fractional oscillator subjected to multiplicative trichotomous noise, Nonlinear Dyn. **90**, (2017) 379-390.
[28] L. Lin, H. Wang, X. Huang, Y. Wen Y, Generalized stochastic resonance for a fractional harmonic oscillator with bias-signal-modulated trichotomous noise, Int. J. Mod. Phys. B 32 (2018) 1850072.
[29] Y. Tian, L. Zhong, G. He, T. Yu, M. Luo, H. E. Stanley, The resonant behavior in the oscillator with double fractional-order damping under the action of nonlinear multiplicative noise, Physica A 490 (2018) 845-856.
[30] R. Ren, K. Deng, Noise and periodic signal induced stochastic resonance in a Langevin equation with random mass and frequency, Physica A 523 (2019) 145-155.
[31] X. Huang, L. Lin, H. Wang, Generalized Stochastic Resonance for a Fractional Noisy Oscillator with Random Mass and Random Damping, J. Stat. Phys. 178 (2020) 1201-1216.
[32] D. Cubero, Finite-size fluctuations and stochastic resonance in globally coupled bistable systems, Phys. Rev. E 77 (2008) 021112.
[33] J. Li, Enhancement and weakening of stochastic resonance for a coupled system, Chaos 21 (2011) 043115.
[34] B. Yang, X. Zhang, L. Zhang, M. Luo, Collective behavior of globally coupled Langevin equations with colored noise in the presence of stochastic resonance, Phys. Rev. E 94 (2016) 022119.
[35] P. Li, R. Ren, Z. Fan, M. Luo, K. Deng, Collective behavior and stochastic resonance in a linear underdamped coupled system with multiplicative dichotomous noise and periodical driving, J. Stat. Mech. (2018) 093206.
[36] S. Zhong, W. Lv, H. Ma, L. Zhang, Collective stochastic resonance behavior in the globally coupled fractional oscillator, Nonlinear Dyn. 94 (2018) 905-923.

[37] L. Lai, L. Zhang, T. Yu, Collective behaviors in globally coupled harmonic oscillators with fluctuating damping coefficient, Nonlinear Dyn. 97 (2019) 2231-2248.

[38] K. M. Hannay, D. B. Forger, V. Booth, Macroscopic models for networks of coupled biological oscillators, Sc. Adv. 4 (2018) e1701047.

[39] B. Cui, A. Zaccone, Generalized Langevin equation and fluctuation-dissipation theorem for particle-bath systems in external oscillating fields, Phys. Rev. E 97 (2018) 060102.

[40] H. Grabert, M. Thorwart, Quantum mechanical response to a driven Caldeira-Leggett bath, Phys. Rev. E 98 (2018) 012122.

[41] V. Hakim, V. Ambegaokar, Quantum theory of a free particle interacting with a linearly dissipative environment, Phys. Rev. A 32 (1985) 423-434.

[42] G. W.Ford, J. T. Lewis, R.F. O' Connell, Quantum Langevin equation, Phys. Rev. A 37 (1988) 4419-4428.

[43] M. Caputo, Linear models of dissipation whose Q is almost frequency independent-II, Geophys. J. R. Astr. Soc. 13 (1967) 529-539.

[44] M. Du, Z. Wang, H. Hu, Measuring memory with the order of fractional derivative, Sci. Reports 3 (2013) 343-345.

[45] K. Wodkiewicz, Theory of nonlinear Langevin equation with quadratic noise, J. Math. Phys. 23 (1982) 2179-2184.

[46] A. Fulinski, Non-Markovian noise, Phys. Rev. E 50 (1994) 2668-2681.

[47] Y. Xu, J. Wu, H. Zhang, S. Ma, Stochastic resonance phenomenon in an underdamped bistable system driven by weak asymmetric dichotomous noise, Nonlinear Dyn. 70 (2012) 531-539.

[48] V. E. Shapiro, V. M. Loginov, "Formulae of differentiation" and their use for solving stochastic equations, Physica A 91 (1978) 563-574.

[49] R. Mankin, A. Ainsaar, E. Reiter, Trichotomous noise-induced transitions, Phys. Rev. E 60 (1999) 1374-1380.

[50] R. Mankin, K. Laas, T. Laas, E. Reiter, Stochastic multiresonance and correlation-time-controlled stability for a harmonic oscillator with fluctuating frequency, Phys. Rev. E 78 (2008) 031120.

[51] T. Yang, H. Zhang, Y. Xu, W. Xu, Stochastic resonance in coupled underdamped bistable systems driven by symmetric trichotomous noises, Intern. J. Nonlinear Mech. 67 (2014) 42-47.

[52] S. Kempfle, I. Schafer, H. Beyer, Fractional calculus via functional calculus: theory and applications. Nonlinear Dyn. 29 (2002) 99-127.

[53] Y. Lin, C. Xu, Finite difference / spectral approximations for the time-fractional diffusion equation, J. Comp. Phys. 225 (2007) 1533-1552.

[54] D. Barik, P. K. Ghosh, D. S. Ray, Langevin dynamics with dichotomous noise; direct simulation and applications, J. Stat. Mech. (2006) P03010.

[55] B. Zhou, D. Lin, Stochastic resonance in a time-delayed bistable system driven by trichotomous noise, Indian J. Phys. 91 (2017) 299-307.